\pdfoutput=1

\documentclass[11pt]{article}

\usepackage{acl}

\usepackage{times}
\usepackage{latexsym}

\usepackage[T1]{fontenc}

\usepackage[utf8]{inputenc}

\usepackage{microtype}

\usepackage{inconsolata}

\usepackage{graphicx}

\usepackage{hyperref}
\usepackage{url}
\usepackage{soul}
\usepackage{pifont}
\usepackage{xcolor}
\usepackage{colortbl}
\usepackage[most]{tcolorbox}
\usepackage{algorithm}
\usepackage{pifont}
\usepackage[noend]{algpseudocode}
\usepackage{amsfonts}
\usepackage{fdsymbol}
\usepackage{graphicx}
\usepackage{subfig}
\usepackage{mathtools}
\usepackage{amsmath}
\usepackage{amsthm}
\usepackage{listings}
\usepackage{wrapfig}
\definecolor{light_gray}{rgb}{.95,.95,.95}
\definecolor{custompurple}{RGB}{93,0,93}
\definecolor{customorange}{RGB}{255,132,6}
\definecolor{customgold}{RGB}{213,177,52}
\definecolor{customblue2}{RGB}{28,205,188}
\definecolor{no_persona_color}{RGB}{152,226,245}
\definecolor{persona_color}{RGB}{193,167,246}
\usepackage{booktabs}
\usepackage{adjustbox}
\usepackage{scalerel}
\usepackage{fontawesome5}

\title{SimUSER: Simulating User Behavior with Large Language Models for Recommender System Evaluation}

\author{
 \textbf{Nicolas Bougie\textsuperscript{1}},
 \textbf{Narimasa Watanabe\textsuperscript{1}}
 \\ \{nicolas.bougie,narimasa.watanabe\}@woven.toyota\\
\\
 \textsuperscript{1}Woven by Toyota,
}

\begin{document}
\maketitle
\begin{abstract}
Recommender systems play a central role in numerous real-life applications, yet evaluating their performance remains a significant challenge due to the gap between offline metrics and online behaviors. Given the scarcity and limits (e.g., privacy issues) of real user data, we introduce SimUSER, an agent framework that serves as believable and cost-effective human proxies. SimUSER first identifies self-consistent personas from historical data, enriching user profiles with unique backgrounds and personalities. Then, central to this evaluation are users equipped with persona, memory, perception, and brain modules, engaging in interactions with the recommender system. SimUSER exhibits closer alignment with genuine humans than prior work, both at micro and macro levels. Additionally, we conduct insightful experiments to explore the effects of thumbnails on click rates, the exposure effect, and the impact of reviews on user engagement. Finally, we refine recommender system parameters based on offline A/B test results, resulting in improved user engagement in the real world.
\end{abstract}

\section{Introduction}\label{sec1}
Recommender systems (RS) have become an indispensable component of our day-to-day lives from e-commerce to social media by offering personalized user experience and improving satisfaction \cite{li2024recent}. Despite their widespread adoption, a key challenge hindering the advancement of the field is evaluation \cite{yoon2024evaluating}. The difficulty arises from the discrepancy between offline metrics (non-interactive), which are typically used during development, and real-life user behaviors, which these systems encounter post-deployment \cite{zhang2019deep}. This results in models that perform well in controlled environments but fail to meet expectations in practical use cases. Such a limitation is further exacerbated by the inherent shortcomings of offline evaluation, notably the inability to measure business values such as user engagement and satisfaction \cite{jannach2019measuring}. On the other hand, online A/B testing is costly to scale up, labor-intensive, and encompasses ethical considerations, underscoring the imperative need for reliable and affordable (interactive) evaluation methods. 

Recent breakthroughs in Large Language Models (LLMs) have shown promise in human behavior modeling by enabling the creation of autonomous agents. In the realm of recommendation systems, RecMind \cite{wang2023recmind} explores the concept of autonomous recommender agents equipped with self-inspiring planning and external tool utilization. Recently, InteRecAgent \cite{huang2023recommender} has extended this idea by proposing memory components, dynamic demonstration-augmented task planning, and reflection. Recently, RecAgent \cite{wang2023user} has attempted to introduce more diverse user behaviors, taking into account external social relationships. Another work, Agent4Rec \cite{hou2024large}, delves into generating faithful user-RS interactions via agent-based simulations, where agents are equipped with a memory module. However, a common characteristic of existing studies is their \textit{insulated nature} --- they primarily rely on knowledge embedded within the model’s weights, neglecting the potential benefits of integrating external knowledge and user-item relationships. Furthermore, prior approaches often disregard user personas and fail to incorporate visual signals, despite their role in shaping user experience and emotion.

To enable synthetic users, we describe an agent architecture built upon LLMs. Our methodology consists of two phases: (1) self-consistent persona matching and (2) recommender system evaluation. In Phase 1, we leverage the semantic awareness of LLMs to extract and identify consistent personas from historical data, encompassing unique backgrounds, personalities, and characteristics. In Phase 2, we impersonate these personas to simulate believable human interactions. This involves a retrieval-augmented framework where the agent interacts with the recommender system based on its persona, memory, perception, and brain modules. The memory module comprises an episodic memory and a knowledge-graph memory. Unlike existing studies that solely rely on text, our perception module incorporates visual cues into the agent's reasoning process. Finally, the brain module is responsible for translating retrieved evidences and graph paths into action plans such as \texttt{[click]}, or \texttt{[exit]}. Following action selection, the user engages in self-reflection to synthesize memories into higher-level inferences and draw conclusions.

\section{Related Work}
Conversational RS initially tackled the recommendation problem using bandit models, emphasizing the quick update of traditional systems through item selection and binary feedback from synthetic users \cite{christakopoulou2016towards}. Taking this further, \cite{zhao2023kuaisim} created a simulation platform where users not only chat about recommendations. Recent techniques have added more natural language flexibility, but user responses are usually limited to binary or multiple-choice formats \cite{lei2020estimation}. In spite of this, these simulations often rely on fixed rules and scripted dialogues, lacking the variability seen in human interactions. To address the above-mentioned limitations, generative simulators using LLMs have been developed, offering more realistic and nuanced conversational abilities \cite{zhang2024agentcf,zhao2023kuaisim}. A few studies have also explored the application of LLMs as recommender systems \cite{hou2024large,li2023gpt4rec,kang2023llms}. These investigations explore LLMs as recommendation engines, rather than as entities that perceive recommendations, thus providing a perspective complementary to our research \cite{wang2024llm,zhang2024prospect}. RecMind \cite{wang2023recmind} proposes self-inspiring agents for recommendation. However, their simulated users are limited to basic actions like rating items, lacking the ability to engage in more complex interactions. Notably, a recent approach \cite{yoon2024evaluating} examines the effectiveness of LLMs as generative users, specifically for conversational recommendation scenarios. A closely related work to ours is Agent4Rec \cite{zhang2023generative} that delves into the generative capabilities of LLMs for modeling user interactions. SimUSER differs significantly from these studies as we utilize detailed personas that are systematically inferred from historical and incorporate a perception module to integrate visual reasoning. Furthermore, SimUSER investigates the potential of graph-based retrieval to represent the rationales underlying user-item interactions. Finally, we introduce multi-round preference elicitation and causal action refinement that leverage retrieved evidences and paths to generate more realistic interactions.

\section{Methodology}
\textbf{Sim}ulated \textbf{USER}s provides a framework for systematically assessing recommender systems by engaging in interactions and providing feedback. Phase 1 matches historical data with a set of personas to enable nuanced and realistic interactions. Phase 2 utilizes the identified personas, historical data, and novel reasoning mechanisms to generate synthetic users with human-like behavior.

\noindent\textbf{Problem Formulation.} Given a user $u \in \mathcal{U}$ and an item $i \in \mathcal{I}$, the aggregated rating of the item is denoted by $R_{i} = \frac{1}{\sum_{u \in \mathcal{U}} y_{ui}} \sum_{u \in \mathcal{U}} y_{ui} \cdot r_{ui}$ where $y_{ui}=0$ indicates that the user $u$ has not rated the item $i$ and inversely $y_{ui}=1$ indicates that the user has rated the item with $r_{ui} \in \{1,2,3,4,5\}$. We also introduce $g_{i} \in G$ as the genre/category of the item. In this study, we seek to discover $y_{ui}$ and $r_{ui}$ for an unseen recommended item $i$.

\subsection{Persona Matching via Consistency Check}
This phase involves assessing the most plausible \textit{persona} based on historical data. A persona $p$ encompasses a set of features that characterize the user: \textbf{age}, \textbf{personality}, and \textbf{occupation}. Personality traits are defined by the Big Five personality facets: \textit{Openness}, \textit{Conscientiousness}, \textit{Extraversion}, \textit{Agreeableness}, and \textit{Neuroticism}, each measured on a scale from 1 to 3. Given the difficulty of obtaining such granular features in real-world settings, our methodology seeks to systematically infer personas from the user's interaction history.

\noindent\textbf{Persona Extraction.} For a user $u$ with interactions $\{(i_{0}, r_{ui_{0}}), \dots, (i_{n}, r_{ui_{n}})\}$, we query the LLM to produce a short summary $s_{u}$ of the user's preferences. To do so, we randomly select 50 items from the user's viewing history. Items rated 4 or above are categorized as \textit{liked}, while those rated below 3 are deemed \textit{disliked}. We then combine $s_{u}$ with historical data to prompt the LLM to generate a persona that matches the interaction history for this user. To enhance the diversity, the LLM is provided a list of possible ages, personalities, and occupations. For each user, a set of $m$ ($m$ = 5) candidate personas is generated, denoted as $\mathcal{P}$.

\noindent\textbf{Self-Consistent Persona Evaluation.} We then assess the consistency of the candidate personas $\mathcal{P}$ to identify the most plausible one. A self-consistency scoring mechanism measures the alignment of candidate personas with historical data. We define a scoring function $s(p,u)$ for each candidate persona $p \in \mathcal{P}$, where $p$ is evaluated against two distinct sets of user-item interactions. For the targeted user $u$, we sample $j$ subsets of $\varrho$ interactions from its history. These are compared with $\varrho$ sampled interactions from other users $\Bar{u}$, denoted as $I_{\Bar{u}}$:
\begin{equation}
    s(p,u) = \sum_{\iota  \sim I_{u}} \hat{r}(\iota,p) - \sum_{\bar{\iota} \in I_{\Bar{u}}} \hat{r}(\bar{\iota},p)
\end{equation}
where $\hat{r}(\iota,p)$ and $\hat{r}(\bar{\iota},p)$ are obtained by querying the LLM to rate the two interaction subsets $\mathcal{\iota}$ and $\bar{\iota}$. Ideally, the LLM agent should assign a higher $\hat{r}(\iota,p)$ for interactions from the targeted user and a lower $\hat{r}(\bar{\iota},p)$ for samples from other users. The candidate persona $p$ with the highest score $s$ is assigned to the user.

\subsection{Engaging in Interactions with RS}

In Phase 2, given a user $u$ and discovered persona $p$, we present a cognitive architecture built upon LLMs comprising four modules: \textbf{persona}, \textbf{perception}, \textbf{memory}, and \textbf{action}.

\subsubsection{Persona Module}
To lay a reliable foundation for the generative agent’s subsequent interactions and evaluations, benchmark datasets are used for initialization of the persona module. An agent’s profile includes the matched persona $p$ along with attributes extracted from its historical data: $p \cup \{\textbf{pickiness}, \textbf{habits}, \textbf{unique tastes}\}$. Since LLMs are biased towards positive sentiment, unless prompted to behave as picky users \cite{yoon2024evaluating}, each agent is assigned a \textit{pickiness} level sampled in \{\textit{not picky}, \textit{moderately picky}, \textit{extremely picky}\} based on the user's average rating. Habits account for user tendencies in engagement, conformity, and variety \cite{zhang2023generative}, while unique tastes are derived from the viewing history summary $s_{u}$ generated in Phase 1.

\subsubsection{Perception Module}
\label{sec:visual_cues}
A primary factor in decision-making is visual stimuli due to their significant influence on curiosity and emotion \cite{liu2024rec}. For instance, when scrolling through a movie recommendation platform, human decisions are heavily driven by the thumbnails of items, which can trigger emotional responses and provide quick visual summaries of the content \cite{koh2022exploration}. To graft these visual elements in an cost-efficient manner, we augment action prompts (see Sec \ref{sec:action_module}) with image-derived captions. The caption $i_{caption}$ of an item $i$ is generated by querying GPT-4o to extract insights that capture emotional tones, visual details, and unique selling points from the item's thumbnail.

\subsubsection{Memory Module}
It is critical for an agent to maintain a memory of the knowledge and experience it has of the world and others. SimUSER uses an episodic memory for interaction history and knowledge-graph memory to capture user-item relationships.

\noindent\textbf{Episodic Memory} stores the interactions with the RS. The memory is initially populated with the user's viewing and rating history. Each time SimUSER executes a new action or rate an item, the corresponding interaction is added to the episodic memory. Drawing from human psychological processes \cite{atkinson1968human}, we use a self-ask retrieval strategy where the LLM generates follow-up questions regarding the query. These questions, along with the initial query, then serve as separate queries for vector similarity search, allowing retrieval of more diverse evidences. For a query $q$, we retrieve top-$k_{1}$ documents using cosine similarity: $s(q,d) = \cos(\mathbf{E}(q), \mathbf{E}(d))$, where $\mathbf{E}$ is an embedding function.

\noindent\textbf{Knowledge-Graph Memory} User behaviors in RS are influenced by both internal factors (personality) and external factors \cite{zhao2014leveraging}. External factors include the influence of others and prior beliefs about items. SimUSER employs a knowledge graph (KG) memory to emulate external influences by retrieving evidences with similar relationships and characteristics. 

\paragraph{Memory Initialization}
The KG memory is initially populated using real-world datasets. It is structured as a graph $\mathcal{G}$, defined as: $\mathcal{G} = \{(h,r,t)| h,t \in \mathcal{V}, r \in \mathcal{E}\}$, in which each triple \textit{(h,r,t)} indicates that a relation $r$ exists from head entity $h$ to tail entity $t$. $\mathcal{V}$ is a set of entities and $\mathcal{E}$ represents relationships between them. For instance, nodes $\mathcal{V}$ may represent entities (e.g., \textit{user}, \textit{item}), while edges $\mathcal{E}$ depict the relationships between these entities (e.g., \textit{liked}). The memory grows with each interaction $i_t$, capturing the evolving nature of user preferences: $\mathcal{G}_{t+1} = \mathcal{G}_t \cup \{(v_i, e_{ij}, v_j) | (v_i, e_{ij}, v_j) \in \mathcal{V} \times \mathcal{E} \times \mathcal{V}\}$.

\noindent\textbf{Graph-Aware Dynamic Item Retrieval} For a user $u$, the retrieval function takes a query item $x$ as input and returns a set of similar items along with their metadata (e.g., \textit{ratings}). We extend PathSim \cite{sun2011pathsim} to capture both user-item and item-item relationships through path-based similarity. A relationship path $p_{x \rightsquigarrow y}$ represents a composite relationship between entities $x$ and $y$ in the form of $x 
 \xrightarrow{\mathcal{E}_{1}} z \xrightarrow{\mathcal{E}_{2}} \dotsc \xrightarrow{\mathcal{E}_{l}} y$, where $\mathcal{E}_{1}$ denotes the edge between entity $x$ and $z$. For example, in the MovieLens network, the co-actor relation can be described using the length-2 relationship path $x \xrightarrow{acts-in} z \xrightarrow{actor} y$. In order to retrieve relevant items based on the query $x$, SimUSER estimates the item-item similarity as:
 \begin{equation}
s_{x, y} = \frac{2 \times \left| \{ p_{x \rightsquigarrow y}: p_{x \rightsquigarrow y} \in \mathcal{P} \} \right|}{\left| p_{x \rightsquigarrow x}: p_{x \rightsquigarrow x} \in \mathcal{P} \right| + \left| p_{y \rightsquigarrow y}: p_{y \rightsquigarrow y} \in \mathcal{P} \right|}
\end{equation}
where $\mathcal{P}$ is the set of paths between query item $x$ and candidate item $y$, and $p_{x \rightsquigarrow y}$ is a path instance. The score $s_{x, y}$ is determined by two factors: (1) the connectivity level, which is the count of paths that connect $x$ and $y$ through $\mathcal{P}$; and (2) the balance of visibility, defined by the number of times these paths are traversed between the two entities. In addition to item-item similarity $s_{x, y}$, we compute user-item similarity $s_{u, y}$ for the target user  $u$ and the candidate item $y$, using the same path-based approach, which is further summed up to $s_{x, y} = \alpha \cdot s_{x, y} + (1 - \alpha) \cdot s_{u, y}$, making retrieval sensitive to both past interactions of the user $u$ and communities in the graph.

\subsection{Brain Module}
\label{sec:action_module}

We endow each agent with a decision-making module that derives subsequent actions. To replicate human-like sequential reasoning, we employ Chain-of-Thought prompting across five key steps.

\textbf{Multi-round Preference Elicitation:} Agents browse items page by page, deciding whether to \texttt{[WATCH]} or \texttt{[SKIP]} based on their preferences and history. To mitigate the inherent positive bias in LLMs, SimUSER incorporates a pickiness modifier (\texttt{You are \{pickiness\} about \{item\_type\}}). When available, we enrich item descriptions with thumbnail captions for multimodal reasoning. A \emph{multi-round} strategy first forms an initial decision $\delta^{(0)}$ based on persona $p$, pickiness $\rho$, and retrieved evidences $E_{k_{1}}$ and $G_{k_{2}}$ from episodic and KG memory. Then, it identifies contradictions between its choice and persona. If conflicts arise or supporting evidence is insufficient, the agent refines its decision: $\delta^{(t)} = \text{LLM}(P_{\text{watch}}, \delta^{(t-1)}, p, E_{k_{1}}, G_{k_{2}})$. To improve decision-making, we expand retrieved documents each round ($k_{1} \leftarrow k_{1} + \Delta_{k}$ and $k_{2} \leftarrow k_{2} + \Delta_{k}$) until reaching a final decision $\delta^{(\text{final})}$.

\textbf{Item Evaluation} After selecting items of interest, agents express both explicit ratings (1-5) and subjective feelings about watched items, which update their memory and influence future cognition. Unlike existing approaches~\cite{zhang2023generative} that neglect rating rationales, Instead, SimUSER leverages the paths of retrieved evidences $i$ from the KG memory, $u \xrightarrow{\mathcal{E}_{1}} z \xrightarrow{\mathcal{E}_{2}} \dotsc \xrightarrow{\mathcal{E}_{l}} i$, They are formatted as plain text and provided as input to the LLM, which generates ratings while explaining how persona, evidences and paths compare to the shortlisted items and influence their rating. 

\textbf{Action Selection:} Based item evaluation and interaction history, agents decide whether to \texttt{[EXIT]} the system, navigate to \texttt{[NEXT]}/\texttt{[PREVIOUS]} pages, or \texttt{[CLICK]} on items for details. This decision involves estimating its satisfaction with previous recommendations, fatigue level, and emotional state. Upon exiting, a satisfaction interview captures opinions about presented recommendations.

\textbf{Causal Action Refinement:} To address suboptimal decision-making (e.g., premature exits), we introduce a \emph{causal reasoning} step where agents generate questions ($Q = LLM(a_{\text{tent}}, H, p, P_{\text{causal}})$) to validate tentative actions. For each counterfactual scenario (e.g., \textit{"What would happen if you exited now?"}), the agent estimates outcomes and adjusts its final action based on cause-effect consistency.

\textbf{Post-interaction Reflection:} Post-interaction reflection lets agents learn from interactions and improve future alignment with their persona. After collecting interaction data, the agent first determines what to reflect on, then extracts insights and cites the particular records that served as evidence for the insights. The post-interaction reflections are fed back into the episodic memory. 

\begin{table*}[tb]
\centering
\begin{adjustbox}{width=\textwidth}
\begin{tabular}{cccccccccccccc}
\toprule
 & \multicolumn{4}{c}{\textbf{MovieLens}} & \multicolumn{4}{c}{\textbf{AmazonBook}} & \multicolumn{4}{c}{\textbf{Steam}} \\ 
\cmidrule(lr){2-5} \cmidrule(lr){6-9} \cmidrule(lr){10-13}
\textbf{Method(1:m)} & \textbf{Accuracy} & \textbf{Precision} & \textbf{Recall} & \textbf{F1 Score} & \textbf{Accuracy} & \textbf{Precision} & \textbf{Recall} & \textbf{F1 Score} & \textbf{Accuracy} & \textbf{Precision} & \textbf{Recall} & \textbf{F1 Score} \\ 
\midrule
RecAgent (1:1) & 0.5807 & 0.6391 & 0.6035 & 0.6205 & 0.6035 & 0.6539 & 0.6636 & 0.6587 & 0.6267 & 0.6514 & 0.6490 & 0.6499 \\
RecAgent (1:3) & 0.5077 & 0.7396 & 0.3987 & 0.5181 & 0.6144 & 0.6676 & 0.4001 & 0.5003 & 0.5873 & 0.6674 & 0.3488 & 0.4576 \\
RecAgent (1:9) & 0.4800 & 0.7491 & 0.2168 & 0.3362 & 0.6222 & 0.6641 & 0.1652 &  0.2647 & 0.5995 & 0.6732 & 0.1744 & 0.2772 \\
\midrule
Agent4Rec (1:1) & 0.6912 & 0.7460 & 0.6914 & 0.6982 & 0.7190 & 0.7276 & 0.7335 & 0.7002 & 0.6892 & 0.7059 & 0.7031 & 0.6786 \\
Agent4Rec (1:3) & 0.6675 & 0.7623 & 0.4210 & 0.5433 & 0.6707 & 0.6909 & 0.4423 & 0.5098 & 0.6505 & 0.7381 & 0.4446 & 0.5194 \\
Agent4Rec (1:9) & 0.6175 & 0.7753 & 0.2139 & 0.3232 & 0.6617 & 0.6939 & 0.2369 & 0.3183 & 0.6021 & 0.7213 & 0.1901 & 0.2822 \\
\midrule
SimUSER (1:1) & \textbf{0.7912} & 0.7976 & \textbf{0.7576} & \textbf{0.7771} & \textbf{0.8221} & \textbf{0.7969} & \textbf{0.7841} & \textbf{0.7904} & \textbf{0.7905} & 0.8033 & \textbf{0.7848} & \textbf{0.7939} \\
SimUSER (1:3) & 0.7737 & 0.8173 & 0.5223 & \textbf{0.6373} & 0.6629 & 0.7547 & 0.5657 & \textbf{0.6467} & 0.7425 & \textbf{0.8048} & 0.5376 & \textbf{0.6446} \\
SimUSER (1:9) & 0.6791 & \textbf{0.8382} & 0.3534 & \textbf{0.4972} & 0.6497 & 0.7588 & 0.3229 & \textbf{0.4530} & 0.7119 & 0.7823 & 0.2675 & \textbf{0.3987} \\
\bottomrule
\end{tabular}
\end{adjustbox}
\caption{User preference alignment across MovieLens, AmazonBook, and Steam datasets.}
\label{table:taste_alignment}
\end{table*}

\section{Experiments}
\textbf{Settings.} All agents are powered by the GPT-4o-mini version of ChatGPT, except when specified differently, with the number of agents set to 1,000. 
\\\textbf{Baselines} We compare SimUSER against RecAgent and Agent4Rec, which represent the closest comparable methods. When possible, we report the results of RecMind, an agent-based RS. Some experiments involve two versions of SimUSER: SimUSER(zero) and SimUSER(sim), where SimUSER(sim) agents first interact with the RS --- grounding interactions and filling their memories, before answering the tasks. 

\subsection{Believably of Synthetic Users}
\label{sec:beliaviability}

In order to appropriately respond to recommendations, synthetic users must possess a clear understanding of their own preferences. Thereby, we query the agents to classify items based on whether their human counterparts have interacted with them or not. We randomly assigned 20 items to each of 1,000 agents, with varying ratios (1:$m$ where $m \in \{1,3,9\}$) of items users had interacted with to non-interacted items ($y_{ui}=0$). We treat this as a binary classification task, taking values between 0 and 1. Table \ref{table:taste_alignment} shows SimUSER agents accurately identified items aligned with their tastes, significantly outperforming RecAgent and Agent4Rec across all distractor levels (paired t-tests, 95\% confidence, $p < 0.002$).

\subsection{Rating Items}
\label{sec:rating}

\begin{table}[tbp]
\centering
\resizebox{1.0\linewidth}{!}{
\begin{tabular}{lcccccc}
\toprule
\textbf{Methods} & \multicolumn{2}{c}{\textbf{MovieLens}} & \multicolumn{2}{c}{\textbf{AmazonBook}} & \multicolumn{2}{c}{\textbf{Steam}} \\
 & \textbf{RMSE} & \textbf{MAE} & \textbf{RMSE} & \textbf{MAE} & \textbf{RMSE} & \textbf{MAE} \\
\midrule
MF & 1.2142 & 0.9971 & 1.2928 & 0.9879 & 1.3148 & 1.0066 \\
AFM & 1.1762 & 0.8723 & 1.3006 & 1.1018 & 1.2763 & 0.9724 \\
RecAgent  & 1.1021 & 0.7632 & 1.2587 & 1.1191 & 1.0766 & 0.9598 \\
RecMind-SI (few-shot) & 1.0651 & 0.6731 & 1.2139 & 0.9434 & 0.9291 & 0.6981 \\
Agent4Rec & 0.7612 & 0.7143 & 0.8788 & 0.6712 & 0.7577 & 0.6880 \\
\midrule
\rowcolor{blue!10}
SimUSER(sim $\cdot$ persona) & \textbf{0.5020} & \textbf{0.4460} & \textbf{0.5676} & \textbf{0.4210} & \textbf{0.5866} & \textbf{0.5323} \\
\phantom{  } SimUSER(zero $\cdot$ w/o persona) & 0.6663 & 0.5501 & 0.6865 & 0.6329 & 0.6976 & 0.6544 \\
\phantom{  } SimUSER(zero $\cdot$ persona)  & 0.5813 & \underline{0.5298} & \underline{0.6542} & \underline{0.5116} & \underline{0.6798} & \underline{0.6151} \\
\phantom{  } SimUSER(sim $\cdot$ w/o persona)  & \underline{0.5844} & 0.5410 & 0.6712 & 0.5441 & 0.6888 & 0.6401 \\
\bottomrule
\end{tabular}
}
\caption{Rating prediction performance. \textbf{Bold}: best results; \underline{underlined}: second-best. SimUSER's improvements are statistically significant ($p< 0.05$).}
\label{fig:rating_prediction}
\end{table}
A key task when interacting with a RS is rating items. We compare several LLM-based baselines, along with traditional recommendation baselines: MF \cite{koren2009matrix} and AFM \cite{xiao2017attentional}. Across all tasks (Table \ref{fig:rating_prediction}), SimUSER considerably outperforms other LLM-powered agents, mainly due to its KG memory that encapsulates priors about items and their relationships with user interactions. Agent4Rec shows higher RMSE due to hallucinations with niche items not embedded in its LLM weights. Notably, incorporating a few steps of simulation always decreases the MAE of the model (SimUSER(sim)). This is because the grounded interactions augment the context during the multi-round assessment, demonstrating that agents can refine their own preferences for unrated items through interactions with the simulator.

\subsection{Recommender System Evaluation}
\begin{table}[tbp]
    \centering
    \resizebox{0.9\linewidth}{!}{
    \begin{tabular}{lcccccc}
    \toprule
     & $\overline{P}_{\text{view}}$ & $\overline{N}_{\text{like}}$ & $\overline{P}_{\text{like}}$ & $\overline{N}_{\text{exit}}$ & $\overline{S}_{\text{sat}}$ \\
    \midrule
    Random & 0.301 & 3.12 & 0.252 & 2.85 & 2.66 \\
    Pop & 0.395 & 4.08 & 0.372 & 2.90 & 3.32 \\
    MF & 0.461 & \textbf{5.91} & 0.443 & 3.05 & 3.65 \\
    MultVAE & \underline{0.514} & 5.38 & \underline{0.455} & \underline{3.18} & \underline{3.78} \\
    LightGCN & \textbf{0.557} & \underline{5.45} & \textbf{0.448} & \textbf{3.29} & \textbf{3.92} \\
    \bottomrule
    \end{tabular}}
    \caption{Evaluation of recommendation strategies on a recommendation task from the MovieLens dataset.}
    \label{tab:results}
\end{table}
Understanding the efficacy of various recommendation algorithms is crucial for enhancing user satisfaction. By simulating human proxies, we can better predict how users will engage with recommender systems, providing valuable interactive metrics. We compare various recommendation strategies, including most popular (Pop), matrix factorization (MF) \cite{koren2009matrix}, LightGCN \cite{he2020lightgcn}, and MultVAE \cite{liang2018variational}, using the MovieLens dataset. Upon exiting, agents rated their satisfaction on a scale from 1 to 10. Ratings above 3 were considered indicative of a \textit{like}. Metrics include average viewing ratio ($\overline{P}_{\text{view}}$), average number of likes ($\overline{N}_{\text{like}}$), average ratio of likes ($\overline{P}_{\text{like}}$), average exit page number ($\overline{N}_{\text{exit}}$), and average user satisfaction score ($\overline{S}_{\text{sat}}$). Table \ref{tab:results} demonstrates that agents exhibit higher satisfaction with advanced recommendations versus random and Pop methods, consistent with real-life trends.

\subsection{LLM Evaluator}

\begin{table}[btp]
\centering
\resizebox{1.0\columnwidth}{!}{
\begin{tabular}{lcccc}
\toprule
 & \textbf{MovieLens} & \textbf{AmazonBook} & \textbf{Steam} \\
\midrule
RecAgent & 3.01 $\pm$ 0.14 & 3.14 $\pm$ 0.13 & 2.96 $\pm$ 0.17   \\
Agent4Rec & 3.04 $\pm$ 0.12 & 3.21 $\pm$ 0.14 & 3.09 $\pm$ 0.16   \\
\rowcolor{blue!10}
SimUSER(w/o persona) & \underline{3.72 $\pm$ 0.18}* & \underline{3.65 $\pm$ 0.21}* & \underline{3.61 $\pm$ 0.24}*  \\
\rowcolor{blue!10}
SimUSER(persona) & \textbf{4.41$\pm$0.16}* & \textbf{3.99$\pm$0.18}* & \textbf{4.02$\pm$0.23}*  \\
\bottomrule
\end{tabular}}
\caption{Human-likeness score evaluated by GPT-4o across recommendation domains. *Significant improvements over best baseline ($p<0.05$).}
\label{tab:llm_evaluator}
\end{table}

As LLM Evaluators \cite{chiang2023can} achieve comparable performance with human evaluators, we use GPT-4o to assess whether agent interactions appear human or AI-generated using a 5-point Likert scale, with higher scores indicating stronger resemblance to human-like responses. Results in Table \ref{tab:llm_evaluator} show our method significantly outperforms Agent4Rec. The memory and persona modules are among the main factors contributing to the faithfulness of our method. We also noticed that letting the agent estimate its tiredness, feeling and emotion greatly enhances the believability and consistency of its responses. On the other hand, Agent4Rec's tendencies to \texttt{[EXIT]} the recommender system early and provide inconsistent ratings for similar items --- ranging from low to high, contribute to suspicions of AI involvement.

\subsection{SimUSER for Offline A/B Testing}

\begin{figure}[tbp]
    \begin{center}
        \includegraphics[width=0.8\linewidth]{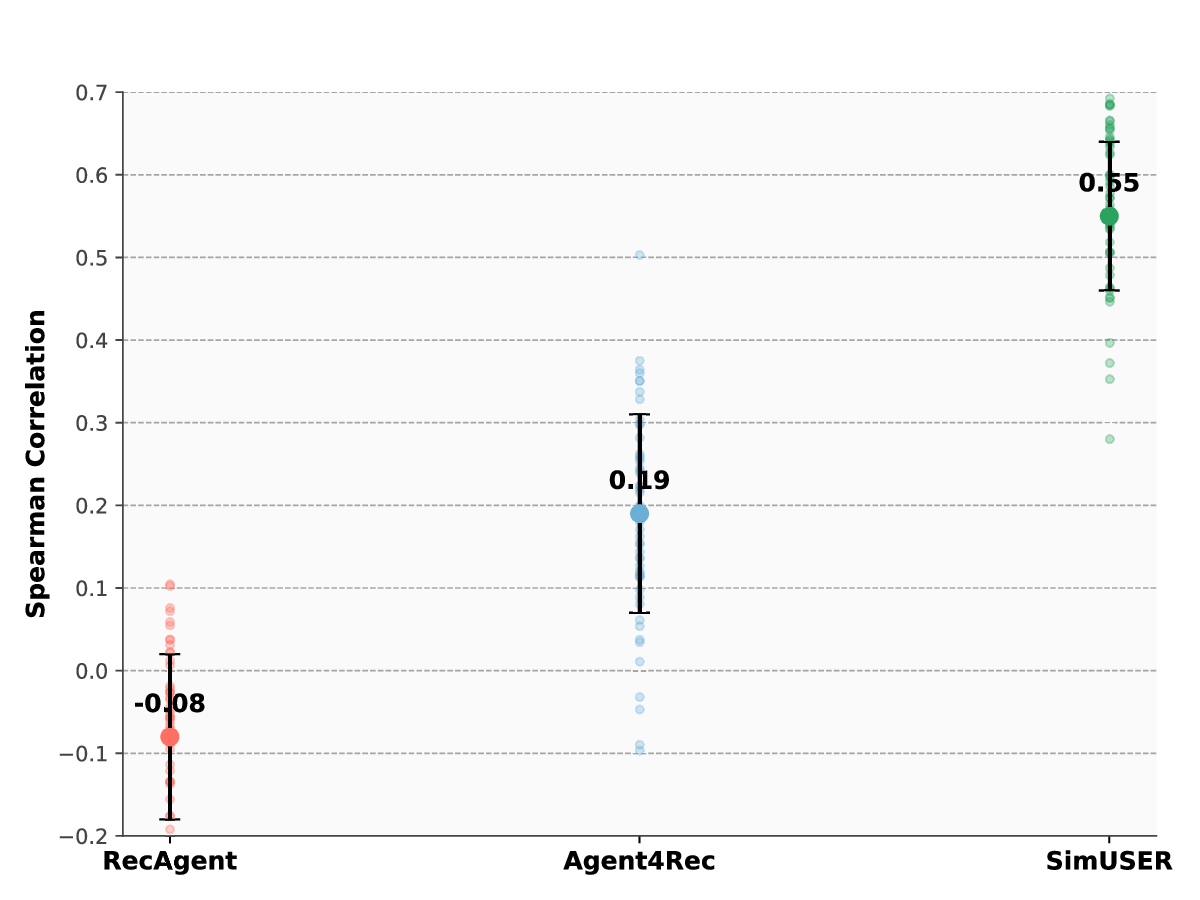} 
    \caption{Spearman correlation between estimated and actual engagement metrics. Higher values indicate better alignment with ground truth metrics.}
    \label{fig:ab_tets}
\end{center}
\end{figure}

We have access a proprietary dataset of 55 online A/B tests, encompassing hundred of thousands of food item recommendations. Each test evaluates variations in recommendation strategies, with the average number of visited pages as the primary business metric. The results, shown in Fig \ref{fig:ab_tets}, indicate that SimUSER achieves the highest correlation with ground truth values, significantly outperforming Agent4Rec and RecAgent. Statistical tests were conducted to validate the significance of SimUSER's performance over the baselines, with p-values below 0.05 for all comparisons. SimUSER effectively captures user engagement, offering a cost-effective alternative to online A/B testing.

\subsection{Optimizing RS with SimUSER}

\begin{table}[tbp]
\centering
\resizebox{1.0\columnwidth}{!}{
\begin{tabular}{lccccc}
\toprule
\textbf{Method} & $\overline{P}_{\text{view}}$ & $\overline{N}_{\text{like}}$ & $\overline{P}_{\text{like}}$ & $\overline{N}_{\text{exit}}$ & $\overline{S}_{\text{sat}}$ \\
\midrule
Baseline & 0.521 & 5.44 & 0.458 & 3.21 & 3.82 \\
Traditional (nDCG@10) & 0.535 & 5.52 & 0.462 & 3.26 & 3.86 \\
SimUSER & \textbf{0.561} & \textbf{5.80} & \textbf{0.517} & \textbf{3.87} & \textbf{4.09} \\
\bottomrule
\end{tabular}}
\caption{Performance comparison of parameter selection strategies on various engagement metrics.}
\label{tab:param_selection_comparison}
\end{table}

We examine whether selecting RS parameters based on SimUSER evaluation or traditional offline metrics (nDCG@10 - \textit{TRAD}), translates to improved business metrics in the real world. We employ the same proprietary dataset. The online performance of the baseline system and the two strategies are presented in Table \ref{tab:param_selection_comparison}. \textit{TRAD} results in performance on par with the original baseline, demonstrating similar findings as in \cite{jannach2019measuring} --- offline metrics do not necessarily translate to business metrics. SimUSER achieves higher engagement and satisfaction, with improvements in average viewing ratio and satisfaction. 

\section{Conclusion}
We present a simulation framework for leveraging LLMs as believable user proxies. Our two-phase approach includes persona matching and interactive RS assessment, seeking to align user interactions more closely with real-world user behaviors. We evaluate SimUSER across various recommendation domains, including movies, books, and video games. Results demonstrate closer alignment of our agents with their human counterparts at both micro and macro levels. We further explore the influence of thumbnails on user engagement and the significance of reviews in user decision-making. Experimental findings highlight the potential of LLM-driven simulations in bridging the gap between offline metrics and business metrics.

\section{Ethics Statement}
This paper proposes an LLM-empowered agent framework designed to simulate user interactions with recommender systems in a realistic and cost-effective manner. While our approach offers significant benefits in terms of scalability and efficiency, it also raises ethical considerations. The use of such agents could lead to unintended consequences, such as bias amplification, where the synthetic agents might inadvertently reinforce existing stereotypes or present skewed recommendations due to biases in the training data. 

Additionally, there is a risk of manipulation of user preferences, as the synthetic agents could be used to subtly influence user behavior by consistently promoting certain types of content without explicit user consent. Furthermore, simulating interactions at a broad scale could result in the identification and exploitation of behavioral patterns that might encourage specific user behaviors, potentially leading to negative societal impacts. Finally, there is a concern that developers or designers might use synthetic users and displace the role of humans and system stakeholders in the design process. We suggest that synthetic uses should not be a substitute for real human input in studies and design processes. Rather, these agents should be leveraged during the initial design phases to explore concepts, especially in situations where recruiting human participants is impractical or where testing certain theories with real people could be challenging or pose risks. By adhering to these principles, we can ensure that the deployment of synthetic users in the wild is ethical and socially responsible.

\bibliography{custom}

\begin{thebibliography}{34}
\providecommand{\natexlab}[1]{#1}

\bibitem[{Atkinson and Shiffrin(1968)}]{atkinson1968human}
Richard~C Atkinson and Richard~M Shiffrin. 1968.
\newblock Human memory: A proposed system and its control processes.
\newblock In \emph{Psychology of learning and motivation}, volume~2, pages 89--195. Elsevier.

\bibitem[{Chiang and Lee(2023)}]{chiang2023can}
Cheng-Han Chiang and Hung-yi Lee. 2023.
\newblock Can large language models be an alternative to human evaluations?
\newblock \emph{arXiv preprint arXiv:2305.01937}.

\bibitem[{Christakopoulou et~al.(2016)Christakopoulou, Radlinski, and Hofmann}]{christakopoulou2016towards}
Konstantina Christakopoulou, Filip Radlinski, and Katja Hofmann. 2016.
\newblock Towards conversational recommender systems.
\newblock In \emph{Proceedings of the 22nd ACM SIGKDD international conference on knowledge discovery and data mining}, pages 815--824.

\bibitem[{F{\"a}rber et~al.(2023)F{\"a}rber, Coutinho, and Yuan}]{farber2023biases}
Michael F{\"a}rber, Melissa Coutinho, and Shuzhou Yuan. 2023.
\newblock Biases in scholarly recommender systems: impact, prevalence, and mitigation.
\newblock \emph{Scientometrics}, 128(5):2703--2736.

\bibitem[{He et~al.(2020)He, Deng, Wang, Li, Zhang, and Wang}]{he2020lightgcn}
Xiangnan He, Kuan Deng, Xiang Wang, Yan Li, Yongdong Zhang, and Meng Wang. 2020.
\newblock Lightgcn: Simplifying and powering graph convolution network for recommendation.
\newblock In \emph{Proceedings of the 43rd International ACM SIGIR conference on research and development in Information Retrieval}, pages 639--648.

\bibitem[{He et~al.(2023)He, Xie, Jha, Steck, Liang, Feng, Majumder, Kallus, and McAuley}]{he2023large}
Zhankui He, Zhouhang Xie, Rahul Jha, Harald Steck, Dawen Liang, Yesu Feng, Bodhisattwa~Prasad Majumder, Nathan Kallus, and Julian McAuley. 2023.
\newblock Large language models as zero-shot conversational recommenders.
\newblock In \emph{Proceedings of the 32nd ACM international conference on information and knowledge management}, pages 720--730.

\bibitem[{Hou et~al.(2024)Hou, Zhang, Lin, Lu, Xie, McAuley, and Zhao}]{hou2024large}
Yupeng Hou, Junjie Zhang, Zihan Lin, Hongyu Lu, Ruobing Xie, Julian McAuley, and Wayne~Xin Zhao. 2024.
\newblock Large language models are zero-shot rankers for recommender systems.
\newblock In \emph{European Conference on Information Retrieval}, pages 364--381. Springer.

\bibitem[{Huang et~al.(2023)Huang, Lian, Lei, Yao, Lian, and Xie}]{huang2023recommender}
Xu~Huang, Jianxun Lian, Yuxuan Lei, Jing Yao, Defu Lian, and Xing Xie. 2023.
\newblock Recommender ai agent: Integrating large language models for interactive recommendations.
\newblock \emph{arXiv preprint arXiv:2308.16505}.

\bibitem[{Jannach and Jugovac(2019)}]{jannach2019measuring}
Dietmar Jannach and Michael Jugovac. 2019.
\newblock Measuring the business value of recommender systems.
\newblock \emph{ACM Transactions on Management Information Systems (TMIS)}, 10(4):1--23.

\bibitem[{Kang et~al.(2023)Kang, Ni, Mehta, Sathiamoorthy, Hong, Chi, and Cheng}]{kang2023llms}
Wang-Cheng Kang, Jianmo Ni, Nikhil Mehta, Maheswaran Sathiamoorthy, Lichan Hong, Ed~Chi, and Derek~Zhiyuan Cheng. 2023.
\newblock Do llms understand user preferences? evaluating llms on user rating prediction.
\newblock \emph{arXiv preprint arXiv:2305.06474}.

\bibitem[{Koh and Cui(2022)}]{koh2022exploration}
Byungwan Koh and Fuquan Cui. 2022.
\newblock An exploration of the relation between the visual attributes of thumbnails and the view-through of videos: The case of branded video content.
\newblock \emph{Decision Support Systems}, 160:113820.

\bibitem[{Koren et~al.(2009)Koren, Bell, and Volinsky}]{koren2009matrix}
Yehuda Koren, Robert Bell, and Chris Volinsky. 2009.
\newblock Matrix factorization techniques for recommender systems.
\newblock \emph{Computer}, 42(8):30--37.

\bibitem[{Lei et~al.(2020)Lei, He, Miao, Wu, Hong, Kan, and Chua}]{lei2020estimation}
Wenqiang Lei, Xiangnan He, Yisong Miao, Qingyun Wu, Richang Hong, Min-Yen Kan, and Tat-Seng Chua. 2020.
\newblock Estimation-action-reflection: Towards deep interaction between conversational and recommender systems.
\newblock In \emph{Proceedings of the 13th International Conference on Web Search and Data Mining}, pages 304--312.

\bibitem[{Li et~al.(2023)Li, Zhang, Wang, Xiong, Lu, and Medioni}]{li2023gpt4rec}
Jinming Li, Wentao Zhang, Tian Wang, Guanglei Xiong, Alan Lu, and Gerard Medioni. 2023.
\newblock Gpt4rec: A generative framework for personalized recommendation and user interests interpretation.
\newblock \emph{arXiv preprint arXiv:2304.03879}.

\bibitem[{Li et~al.(2024)Li, Liu, Satapathy, Wang, and Cambria}]{li2024recent}
Yang Li, Kangbo Liu, Ranjan Satapathy, Suhang Wang, and Erik Cambria. 2024.
\newblock Recent developments in recommender systems: A survey.
\newblock \emph{IEEE Computational Intelligence Magazine}, 19(2):78--95.

\bibitem[{Liang et~al.(2018)Liang, Krishnan, Hoffman, and Jebara}]{liang2018variational}
Dawen Liang, Rahul~G Krishnan, Matthew~D Hoffman, and Tony Jebara. 2018.
\newblock Variational autoencoders for collaborative filtering.
\newblock In \emph{Proceedings of the 2018 world wide web conference}, pages 689--698.

\bibitem[{Liu et~al.(2024)Liu, Wang, Sun, and Yu}]{liu2024rec}
Yuqing Liu, Yu~Wang, Lichao Sun, and Philip~S Yu. 2024.
\newblock Rec-gpt4v: Multimodal recommendation with large vision-language models.
\newblock \emph{arXiv preprint arXiv:2402.08670}.

\bibitem[{Maas et~al.(2011)Maas, Daly, Pham, Huang, Ng, and Potts}]{maas2011learning}
Andrew Maas, Raymond~E Daly, Peter~T Pham, Dan Huang, Andrew~Y Ng, and Christopher Potts. 2011.
\newblock Learning word vectors for sentiment analysis.
\newblock In \emph{Proceedings of the 49th annual meeting of the association for computational linguistics: Human language technologies}, pages 142--150.

\bibitem[{Nguyen et~al.(2018)Nguyen, Maxwell~Harper, Terveen, and Konstan}]{nguyen2018user}
Tien~T Nguyen, F~Maxwell~Harper, Loren Terveen, and Joseph~A Konstan. 2018.
\newblock User personality and user satisfaction with recommender systems.
\newblock \emph{Information Systems Frontiers}, 20:1173--1189.

\bibitem[{Sun et~al.(2011)Sun, Han, Yan, Yu, and Wu}]{sun2011pathsim}
Yizhou Sun, Jiawei Han, Xifeng Yan, Philip~S Yu, and Tianyi Wu. 2011.
\newblock Pathsim: Meta path-based top-k similarity search in heterogeneous information networks.
\newblock \emph{Proceedings of the VLDB Endowment}, 4(11):992--1003.

\bibitem[{Wang et~al.(2023{\natexlab{a}})Wang, Zhang, Yang, Chen, Tang, Zhang, Chen, Lin, Song, Zhao et~al.}]{wang2023user}
Lei Wang, Jingsen Zhang, Hao Yang, Zhiyuan Chen, Jiakai Tang, Zeyu Zhang, Xu~Chen, Yankai Lin, Ruihua Song, Wayne~Xin Zhao, and 1 others. 2023{\natexlab{a}}.
\newblock User behavior simulation with large language model based agents.
\newblock \emph{arXiv preprint arXiv:2306.02552}.

\bibitem[{Wang et~al.(2024)Wang, Wu, Hong, Liu, and Fu}]{wang2024llm}
Xinyuan Wang, Liang Wu, Liangjie Hong, Hao Liu, and Yanjie Fu. 2024.
\newblock Llm-enhanced user-item interactions: Leveraging edge information for optimized recommendations.
\newblock \emph{arXiv preprint arXiv:2402.09617}.

\bibitem[{Wang et~al.(2023{\natexlab{b}})Wang, Jiang, Chen, Yang, Zhou, Cho, Fan, Huang, Lu, and Yang}]{wang2023recmind}
Yancheng Wang, Ziyan Jiang, Zheng Chen, Fan Yang, Yingxue Zhou, Eunah Cho, Xing Fan, Xiaojiang Huang, Yanbin Lu, and Yingzhen Yang. 2023{\natexlab{b}}.
\newblock Recmind: Large language model powered agent for recommendation.
\newblock \emph{arXiv preprint arXiv:2308.14296}.

\bibitem[{Xiao et~al.(2017)Xiao, Ye, He, Zhang, Wu, and Chua}]{xiao2017attentional}
Jun Xiao, Hao Ye, Xiangnan He, Hanwang Zhang, Fei Wu, and Tat-Seng Chua. 2017.
\newblock Attentional factorization machines: Learning the weight of feature interactions via attention networks.
\newblock \emph{arXiv preprint arXiv:1708.04617}.

\bibitem[{Yang et~al.(2023)Yang, Zhang, and Li}]{YangZL23}
Heng Yang, Chen Zhang, and Ke~Li. 2023.
\newblock Pyabsa: {A} modularized framework for reproducible aspect-based sentiment analysis.
\newblock pages 5117--5122.

\bibitem[{Yao et~al.(2024)Yao, Yu, Zhang, Wang, Cui, Zhu, Cai, Li, Zhao, He, Chen, Zhou, Zou, Zhang, Hu, Zheng, Zhou, Cai, Han, Zeng, Li, Liu, and Sun}]{yao2024minicpmv}
Yuan Yao, Tianyu Yu, Ao~Zhang, Chongyi Wang, Junbo Cui, Hongji Zhu, Tianchi Cai, Haoyu Li, Weilin Zhao, Zhihui He, Qianyu Chen, Huarong Zhou, Zhensheng Zou, Haoye Zhang, Shengding Hu, Zhi Zheng, Jie Zhou, Jie Cai, Xu~Han, and 4 others. 2024.
\newblock Minicpm-v: A gpt-4v level mllm on your phone.
\newblock \emph{arXiv preprint 2408.01800}.

\bibitem[{Yoon et~al.(2024)Yoon, He, Echterhoff, and McAuley}]{yoon2024evaluating}
Se-eun Yoon, Zhankui He, Jessica~Maria Echterhoff, and Julian McAuley. 2024.
\newblock Evaluating large language models as generative user simulators for conversational recommendation.
\newblock \emph{arXiv preprint arXiv:2403.09738}.

\bibitem[{Yuan et~al.(2022)Yuan, Yuan, Li, Kong, Li, Chen, Yang, Yu, Hu, Li, Xu, and Qie}]{yuan2022tenrec}
Guanghu Yuan, Fajie Yuan, Yudong Li, Beibei Kong, Shujie Li, Lei Chen, Min Yang, Chenyun Yu, Bo~Hu, Zang Li, Yu~Xu, and Xiaohu Qie. 2022.
\newblock \href {https://openreview.net/forum?id=PfuW84q25y9} {Tenrec: A large-scale multipurpose benchmark dataset for recommender systems}.
\newblock In \emph{Thirty-sixth Conference on Neural Information Processing Systems Datasets and Benchmarks Track}.

\bibitem[{Zhang et~al.(2023)Zhang, Sheng, Chen, Li, Deng, Wang, and Chua}]{zhang2023generative}
An~Zhang, Leheng Sheng, Yuxin Chen, Hao Li, Yang Deng, Xiang Wang, and Tat-Seng Chua. 2023.
\newblock On generative agents in recommendation.
\newblock \emph{arXiv preprint arXiv:2310.10108}.

\bibitem[{Zhang et~al.(2024{\natexlab{a}})Zhang, Bao, Wang, Zhang, Shi, Xu, Feng, and Chua}]{zhang2024prospect}
Jizhi Zhang, Keqin Bao, Wenjie Wang, Yang Zhang, Wentao Shi, Wanhong Xu, Fuli Feng, and Tat-Seng Chua. 2024{\natexlab{a}}.
\newblock Prospect personalized recommendation on large language model-based agent platform.
\newblock \emph{arXiv preprint arXiv:2402.18240}.

\bibitem[{Zhang et~al.(2024{\natexlab{b}})Zhang, Hou, Xie, Sun, McAuley, Zhao, Lin, and Wen}]{zhang2024agentcf}
Junjie Zhang, Yupeng Hou, Ruobing Xie, Wenqi Sun, Julian McAuley, Wayne~Xin Zhao, Leyu Lin, and Ji-Rong Wen. 2024{\natexlab{b}}.
\newblock Agentcf: Collaborative learning with autonomous language agents for recommender systems.
\newblock In \emph{Proceedings of the ACM on Web Conference 2024}, pages 3679--3689.

\bibitem[{Zhang et~al.(2019)Zhang, Yao, Sun, and Tay}]{zhang2019deep}
Shuai Zhang, Lina Yao, Aixin Sun, and Yi~Tay. 2019.
\newblock Deep learning based recommender system: A survey and new perspectives.
\newblock \emph{ACM computing surveys (CSUR)}, 52(1):1--38.

\bibitem[{Zhao et~al.(2023)Zhao, Liu, Cai, Zhao, Liu, Zheng, Jiang, and Gai}]{zhao2023kuaisim}
Kesen Zhao, Shuchang Liu, Qingpeng Cai, Xiangyu Zhao, Ziru Liu, Dong Zheng, Peng Jiang, and Kun Gai. 2023.
\newblock Kuaisim: A comprehensive simulator for recommender systems.
\newblock \emph{Advances in Neural Information Processing Systems}, 36:44880--44897.

\bibitem[{Zhao et~al.(2014)Zhao, McAuley, and King}]{zhao2014leveraging}
Tong Zhao, Julian McAuley, and Irwin King. 2014.
\newblock Leveraging social connections to improve personalized ranking for collaborative filtering.
\newblock In \emph{Proceedings of the 23rd ACM international conference on conference on information and knowledge management}, pages 261--270.

\end{thebibliography}
\clearpage
\appendix

\sethlcolor{customblue}

\begin{figure*}[tbp]
    \begin{center}
        \includegraphics[width=0.7\linewidth]{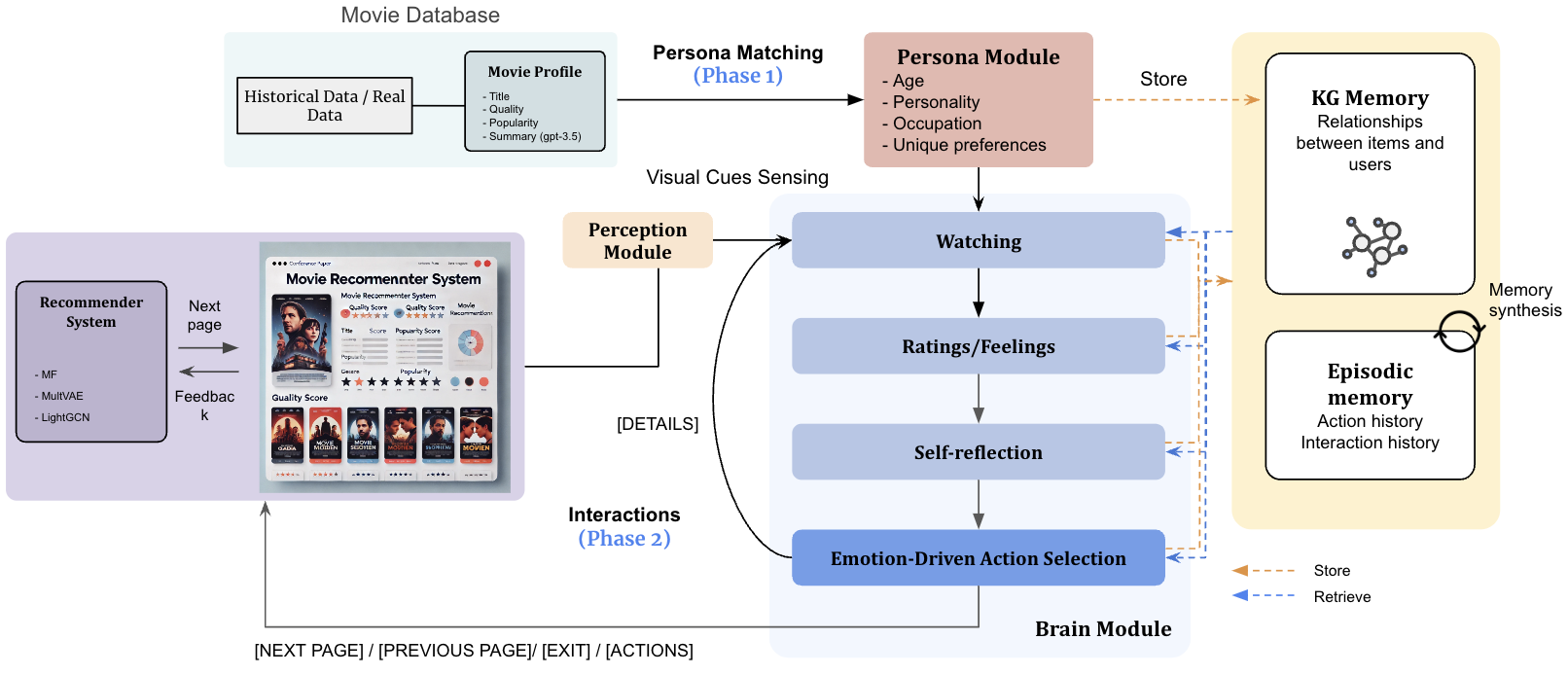}
    \caption{The SimUSER framework for evaluating a movie recommender system.}
    \label{fig:overall_method}
\end{center}
\end{figure*}

\section{Experimental Setup}
\textbf{Experimental Settings.} We separate the dataset into training, validation, and test sets (80/10/10\%), using a time-based split. This ensures to reflect temporal distribution shift that may be observed in the real-world. Relationships between users and items from the training/validation and test sets were excluded from the knowledge graph memory to prevent data leakage. These datasets are employed for the initialization of each agent --- persona and memory modules, as well as self-consistent persona matching. In order to address privacy concerns, the name and gender are omitted. Moreover, for the sake of generality, we do not utilize user-specific information available in these datasets, relying instead on the personas identified in Phase 1 of SimUSER.

In this paper, we report results for SimUSER with simulation \textbf{SimUSER(sim)}, and without simulation \textbf{SimUSER(zero)}. In SimUSER(zero), the agent's memory module is initialized from the history of its human counterpart. When the review score for an item is greater than 4, the agent stores a memory entry in the form \texttt{I liked \{item\_name\} based on my review score of \{score\}}. For a score of 2 or below, the following format is utilized \texttt{I disliked \{item\_name\} based on my review score of \{score\}}. Neutral scores result in the entry \texttt{I felt neutral about \{item\_name\} based on my review score of \{score\}}. In SimUSER(sim), agents can also interact with the recommender systems (training set) for up to 20 pages or exit the system at any time. The corresponding interactions are used to enhance the memory module. In all the experiments, items rated $\geq 4$ are considered as liked by the user, while items $\leq$ 2 are considered as disliked. These interactions are stored both as plain text in the episodic memory and as relationships in the knowledge graph memory. These simulated interactions with the RS are stored in the episodic memory with the following format: \texttt{The recommender system recommended the following \{item\_type\} to me on page \{page\_number\}: \{name\_all\_items\}, among them, I selected \{watched\_items\} and rate them \{ratings\} respectively. I dislike the rest \{item\_type\} items: \{dislike\_items\}}.

\noindent In some sets of experiments, we report performance without persona matching SimUSER(w/o persona), and with persona matching SimUSER(persona). In the absence of persona matching, personality traits, age, occupation and taste summary are omitted from the prompts. Matrix factorization (MF) is utilized as the recommender model unless specified otherwise. In our simulator, agents are presented with four items $n=4$ per page and allowed to interact by viewing and rating items based on their preferences. When the output of the LLM deviated from the desired format, resulting in errors, the LLM was re-prompted with the following instruction: \texttt{You have one more chance to provide the correct answer}.

The path-score used during the retrieval of evidences from the KG memory, we further combine this score with user-item similarity ($s_{x, y} = \alpha \cdot s_{x, y} + (1 - \alpha) \cdot s_{u, y}$) and enhance it with semantic similarity using embeddings from OpenAI's \textit{text-embedding-3-small} model. The top-$k_{2}$ items, their attributes, and paths are returned to condition the brain module.

As mentioned above, we leverage \texttt{GPT-4o-mini} as the LLM backbone in all the experiments unless stated differently. We use $\alpha=0.8$ to balance item-item similarity with user-item similarity. We set $k_{2}=3$ when retrieving similar items from the knowledge graph-memory, and $k_{1}=5$ for the episodic memory. The titles and ratings of retrieved items from the knowledge graph are concatenated to condition decision-making prompts. Empirically, we set the weight of node embeddings to 0.25 when computing top-$k_{2}$ scores. Documents and embedding of text ($\mathbf{E}$) were obtained using \texttt{text-embedding-3-small}. Given the average rating $\bar{R}$ of a user: $ \bar{R} = \frac{1}{N} \sum_{i=1}^{N} r_{ui}$, the pickiness level \( P(\bar{R}) \) of a user was determined based on the following thresholds:\[
P(\bar{R}) = 
\begin{cases} 
P_1 & \text{if } \bar{R} \geq 4.5 \\
P_2 & \text{if } 3.5 \leq \bar{R} < 4.5 \\
P_3 & \text{if } \bar{R} < 3.5 
\end{cases}
\]
where \( P_1 \) corresponds to \textit{not picky}, \( P_2 \) corresponds to \textit{moderately picky}, and \( P_3 \) corresponds to \textit{extremely picky}. 

The persona attributes are estimated as follows:
\begin{itemize}
    \item Engagement quantifies the frequency and breadth of a user’s interactions with recommended items, distinguishing between users who extensively watch and rate many of items and those who confine themselves to a minimal set. The engagement trait for user $u$ can be mathematically expressed as: $T^{u}_{act} = \sum_{i\in \mathcal{I}} y_{ui}$.
    \item Conformity measures how closely a user’s ratings align with average item ratings, drawing a distinction between users with unique perspectives and those whose opinions closely mirror popular sentiments. For user $u$, the conformity trait is defined as: $T_{conf}^{u} = \frac{1}{\sum_{i\in \mathcal{I}} y_{ui}} \sum_{i\in \mathcal{I}} y_{ui} \cdot |r_{ui} - R_{i}|^{2}$.
    \item Variety reflects the user’s proclivity toward a diverse range of item genres or their inclination toward specific genres. The variety trait for user $u$ is formulated as: $T_{div}^{u} = |U_{i\in \{y_{ui}=1\}}g_{i}|$. To encode users’ unique tastes in natural language, we utilize the summary $s_{u}$ obtained in Phase 1, which describes long-term preferences. 
\end{itemize} 

To generate captions, for each item $i$, we first generate an initial caption draft $i^{*}$ by querying: $i^{*} = LLM\bigl(P_{\text{caption}},\,i\bigr)$, where $P_{\text{caption}}$ is the task prompt. To reduce hallucination, we then decompose $i^{*}$ into atomic claims $\{\,a_1,\dots,a_m\}$, each describing a specific, factual statement (e.g., ``The movie is scary''), rather than subjective opinions. Next, each claim $a_k$ is formed into a polar (yes/no) query, and an open-source MLLM \cite{yao2024minicpmv} is queried to generate the confidence of agreement and disagreement as the claim score $s_{a} = (p_{yes}, p_{no})$, where $p_{yes}$ is the probability of answering with ``yes'' and $p_{no}$ is the probability of answering with ``no''. Finally, the original caption is refined in order to obtain a the item's caption $i_{caption} = LLM(i^{*}, P_{\text{combine}}, (a, s_{a}), ...)$. This minimizes the risk of agents selecting items based on inaccurate captions by ensuring the generated descriptions are fact-based and supported by confidence scores.

In Appendix \ref{sec:eff_images}, we compare the results of SimUSER taking as input: 1) the original movie poster, 2) a random screenshot from the movie trailer on YouTube, 3) the original movie poster distorted with a blue color filter (hue=30, lightness=30, saturation=30). An illustration of the method is provided in Figure \ref{fig:overall_method}, detailing the interaction between its components and their roles within the proposed framework.

\subsection{Brain Module Details}
\label{sec:action_module}
We now provide a comprehensive explanation of the Brain Module, detailing the implementation and technical details. To replicate human-like sequential reasoning, we employ Chain-of-Thought prompting, repeatedly performing the five steps.

\subsection{Multi-round Preference Elicitation}
We employ a \emph{multi-round} preference elicitation strategy to refine the user’s choice. First, an initial decision \( \delta^{(0)} \) is formed based on the agent’s persona \( p \), pickiness level $\rho$, and retrieved evidences $E_{k_{1}}$ $G_{k_{2}}$ from the episodic and KG memory respectively. Along with this decision, the agent provides a reason for its choice and cites the supporting evidence, if any. Next, the agent checks for contradictions, such as deciding to watch a pure horror film while the persona indicates strong aversion to horror. If a conflict arises or cannot find enough supporting evidences, the agent is prompted to confirm or modify the initial decision, resulting in an updated decision $\delta^{(t)}  = LLM(P_{\text{watch}}, \delta^{(t-1)}, p, E_{k_{1}}, G_{k_{2}})$, where $P_{watch}$ is the task prompt, and $\mathbf{G}_{t}$ and $\mathbf{E}_{(t)}$ are retrieved evidences. To assist the agent’s decision-making, we \emph{expand} the retrieved documents at each round: \( k_{1} \leftarrow k_{1} + \Delta_{k} \) and \( k_{2} \leftarrow k_{2} + \Delta_{k} \), exposing additional relevant items or past interactions. This continues until a final decision $\delta^{(\text{final})}$ is reached. 

\subsection{Providing Feelings and Rating Items}

Once the user identifies the items of interest $\delta^{(\text{final})} = \{i_{1}, ...\}$, they express their reactions through both explicit ratings and subjective feelings. Intuitively, a real user may produce much feelings after watching an item, which will be stored in their memory and influence their future cognition and behaviors. Along with the item rating $\in \{1,2,3,4,5\}$, we query the user's feelings about the watched items and leverage such information to update the memory module. Newly liked and disliked items are fed back into the memory module. Existing approaches \cite{zhang2023generative} neglect the underlying rationale behind user ratings. Instead, SimUSER leverages the paths of each retrieved evidences $i$ from the KG memory, $u \xrightarrow{\mathcal{E}_{1}} z \xrightarrow{\mathcal{E}_{2}} \dotsc \xrightarrow{\mathcal{E}_{l}} i$. These paths are formatted as plain text and provided as input to the LLM, which generates ratings while explaining how persona, evidences and paths compare to the shortlisted items and influence their rating.

\subsection{Emotion-driven Action Selection}

The agent decides ($a_{\mathrm{tent}}$) whether to \texttt{[EXIT]} the system, go to \texttt{[NEXT]} page, return to a \texttt{[PREVIOUS]} page, or \texttt{[CLICK]} on an item to access more details. If the agent decides to click on an item, the item is displayed with an extended description that replaces the short $\{item\_description\}$, which is then used to determine whether it wishes to engage further with the item. Finally, if \texttt{[EXIT]} is selected, a satisfaction interview is conducted to gather granular opinions and ratings on the presented recommendations. To this end, the agent sequentially: 1) estimates its satisfaction level with preceding recommendations, 2) generates its current fatigue level \cite{zhang2023generative}, 3) infers its current emotion, such as \texttt{EXCITED}, and 4) selects the most suitable action. Satisfaction level, fatigue, and emotion are dynamic elements that the agent employs to adapt its actionable plan with the recommender system.

\subsection{Causal Action Refinement}

Suboptimal decision-making (e.g., premature exits or misaligned clicks) can arise as the agent struggles to understand the impact of its decision, necessitating iterative adjustments to align with implicit preferences. In light of this, we introduce a \textit{causal} reasoning step which encourages the assistant to actively seek to understand the causal relationships between its decisions and latent user-state dynamics. Assuming the tentative action $a_{\mathrm{tent}}$ and context $H$, the LLM generates causal questions $Q$ to validate the rationale behind $a_{\mathrm{tent}}$, $Q = LLM(a_{\text{tent}}, H, p, P_{\text{causal}})$, where $P_{\text{causal}}$ refers to a predetermined prompt. Causal questions may for example be: \textit{Does tiredness reduce the appeal of this action?}, \textit{What would happen if you exited the system now?}. For each counterfactual, the LLM estimates outcomes such as satisfaction, alignment with persona, and fatigue. This includes a scalar $s_{q}$ and textual verdict $v_{q}$ reflecting how \emph{cause-effect} relationships support or contradict $a_{\mathrm{tent}}$. Finally, the LLM is queried to adjust the action if the consistency score is low, $a_{\text{final}} = LLM(a_{\text{tent}}, H, p, P_{\text{action}}, \Pi_{q \in Q} \{q, s_{q},  v_{q}\})$.

\section{Pseudo-Code}
We present the pseudo-code for SimUSER agent. 
\begin{algorithm}[H]
\caption{SimUSER Algorithm}
\begin{algorithmic}[1]
\State \textbf{Input:} Historical data $H_u$ for user $u$
\State \textbf{Output:} Simulated interactions and feedback

\State \textbf{Phase 1: Persona Matching}
\State $\mathcal{P} \leftarrow$ Generate persona from $H_u$
\State $p \leftarrow$ Identify best persona $\in \mathcal{P}$ using self-consistency score

\State \textbf{Phase 2: Simulate Interactions}
\State Initialize memory module from $H_u$
\Repeat
    \State Perceive the page and items \Comment{Generate captions} 
    \State Retrieve similar items from the KG memory
    \State Decide what items to watch
    \State Rate the items and provide feelings
    \State Decide next action $a$ based on satisfaction, fatigue, and emotion
    \State Perform post-interaction reflection
    \State Update memory module
    \If {$a = \texttt{[EXIT]}$}
        \State \textbf{break}
    \Else
    \State Perform action $a$
    \EndIf
\Until{Maximum number of pages reached}

\State \textbf{Return} Simulated interactions, metrics, and feedback
\end{algorithmic}
\end{algorithm}

\section{Additional Experiments}

\subsection{Rating Distribution}

\begin{figure}[tbp]
    \centering
    \includegraphics[width=0.90\linewidth]{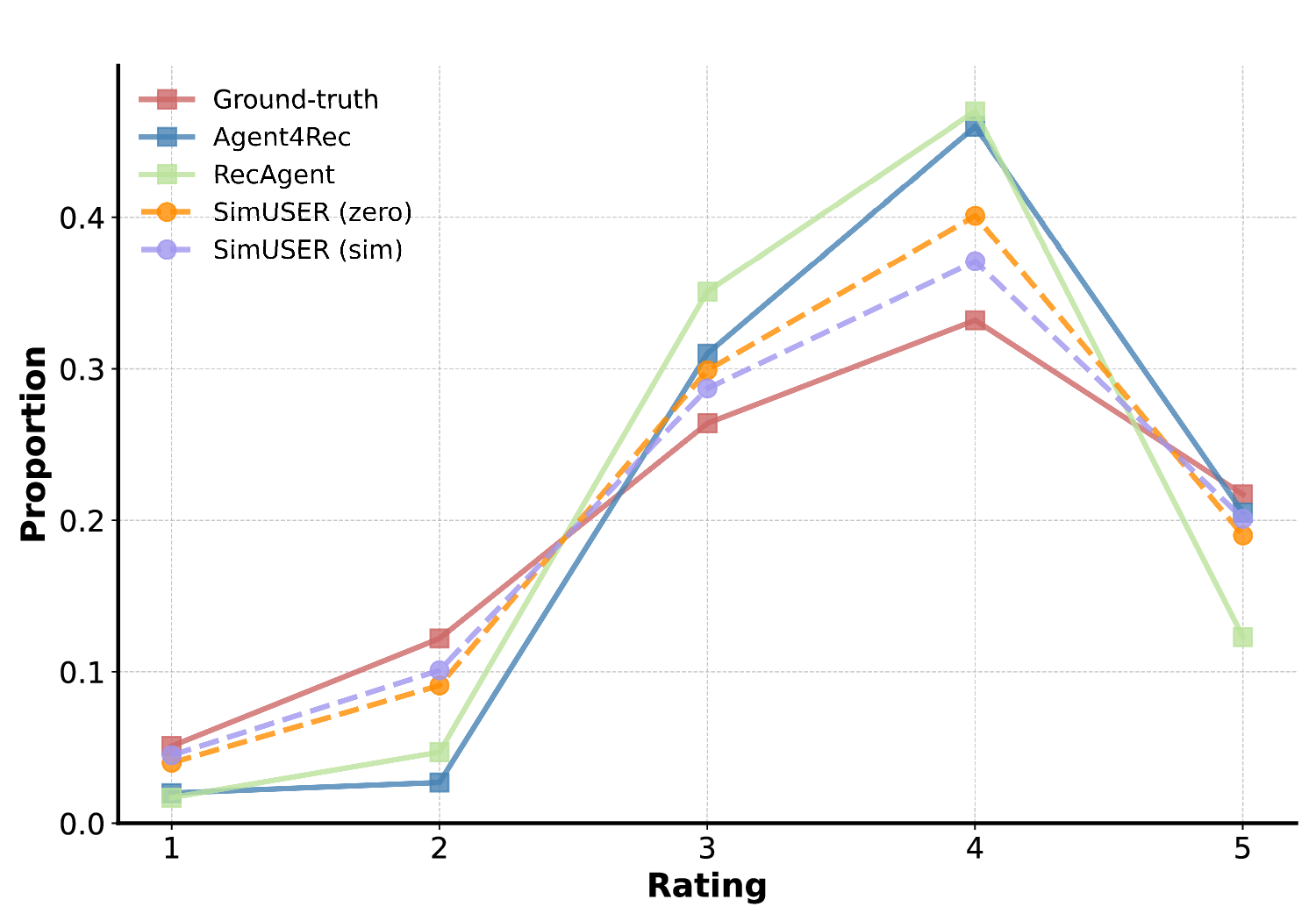} 
    \caption{Comparison of rating distributions between ground-truth and human proxies.}
    \label{fig:rating_alignement}
\end{figure}

Beyond individual rating alignment, human proxies must replicate real-world behavior at the macro level. This implies ensuring that the distribution of ratings generated by the agents aligns closely with the distributions observed in the original dataset. Figure \ref{fig:rating_alignement} presents the rating distribution from the MovieLens-1M dataset and the ratings generated by the agents. These results reveal a high degree of alignment between the simulated and actual rating distributions, with a predominant number of ratings at 4 and a small number of low ratings (1-2). While Agent4Rec assigns fewer 1-2 ratings than real users, our approach, by retrieving past interactions from the episodic memory, allows agents to contextualize their ratings based on a broader and more consistent understanding of their own preferences.

\subsection{Alignment: Rating vs Feeling}

\begin{figure}[tbp]
    \begin{center}
        \includegraphics[width=1.0\linewidth]{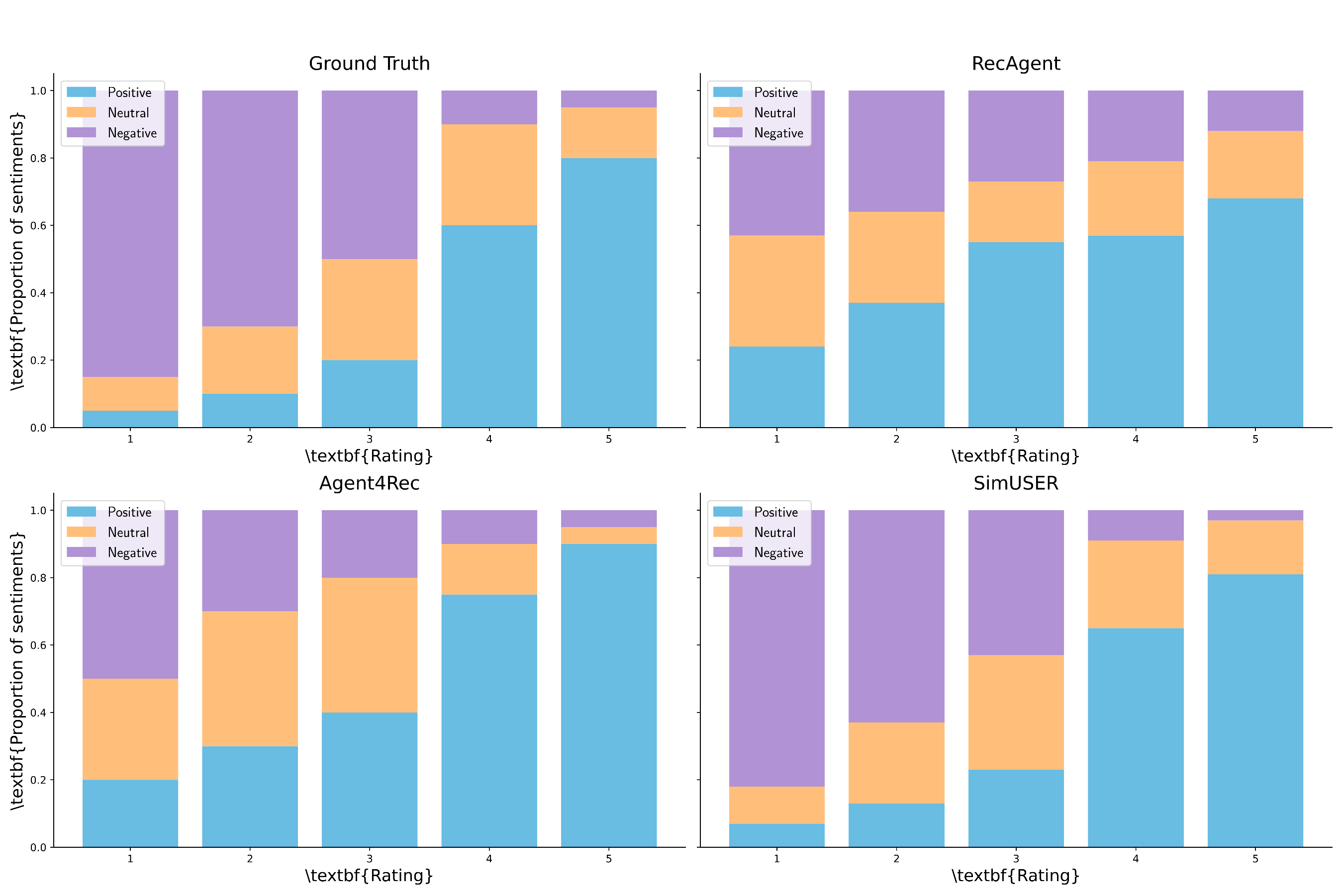} 
    \caption{Ratings vs feelings on IMDB dataset. Comparison between human (top left) and LLM-empowered agents.}
    \label{fig:rating_aligmenet}
\end{center}
\end{figure}
Expressing aligned reviews and ratings is of primary importance to simulate realistic human proxies. Thus, in this section we delve into the alignment between ratings and sentiments. In detail, we prompt the agent to assume one has interacted with a certain item, and ask about its rating and feelings on it. Reviews and ratings from IMDB \cite{maas2011learning} are used as ground truth since MovieLens does not contain reviews. After getting a collection of responses, we conduct sentiment-based analysis with PyABSA \cite{YangZL23}. We compare the rating and sentiment distributions of: humans, RecAgent, Agent4Rec, and SimUSER. As depicted in Figure \ref{fig:rating_aligmenet}, our agents generate ratings aligned with their opinions. For instance, ratings $\geq 4$ are typically associated with positive sentiments. In contrast, Agent4Rec exhibits a bias towards positive opinions, resulting in more positive feelings about the items, including when generating low ratings. It is noteworthy that SimUSER agents and genuine humans express similar sentiments at a macro level.

\subsection{Preference Coherence}

\begin{figure}[tbp]
    \begin{center}
        \includegraphics[width=1.0\linewidth]{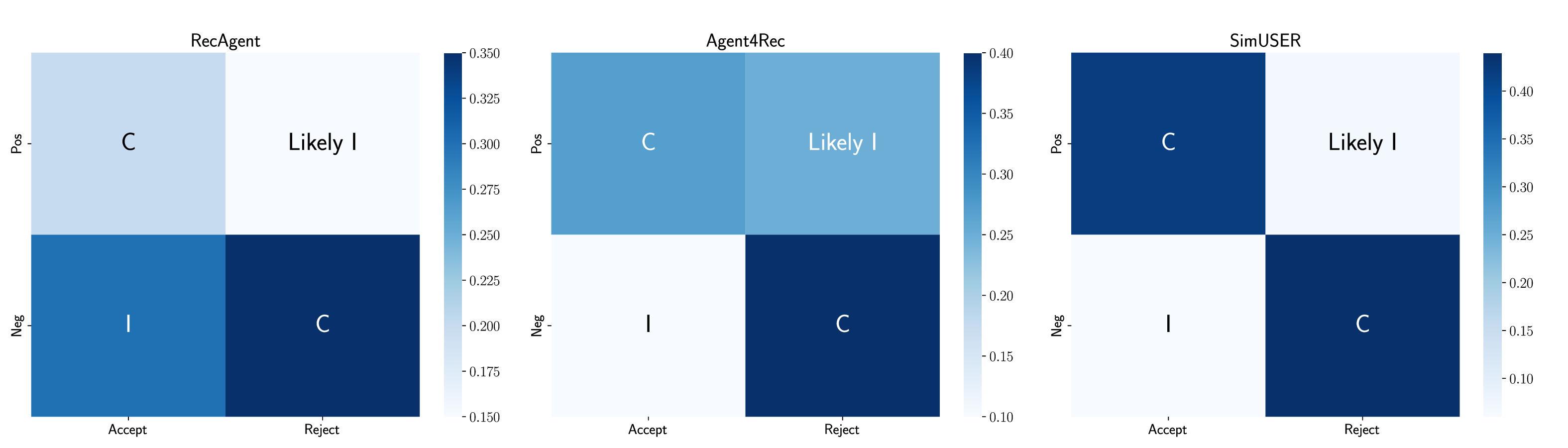} 
    \caption{Preference coherence (accept/reject task). 'I' stands for incoherent; 'C' stands for coherent (Reddit dataset). }
    \label{fig:coherence_watch}
\end{center}
\end{figure}

Under this scenario, we aim to evaluate whether agents prefer positive recommendations based on a query. Namely, for each request in the Reddit dataset \cite{he2023large}, we sample: (1) a comment from this request (positive recommendation) (2) a random comment (negative recommendation). The agent is then prompted to decide which items to \textit{watch}. Ideally, synthetic users should watch only positive recommendations and decline negative ones. Behavior is incoherent when the simulator accepts a negative recommendation. We clearly see in Figure \ref{fig:coherence_watch} that our agents are overall coherent, but sometimes prefer negative recommendations, its proportion being around 4\%. Particularly, Agent4Rec agents often accept recommendations that are not aligned with their age and personality.

\begin{figure}[tbp]
    \begin{center}
        \includegraphics[width=1.0\linewidth]{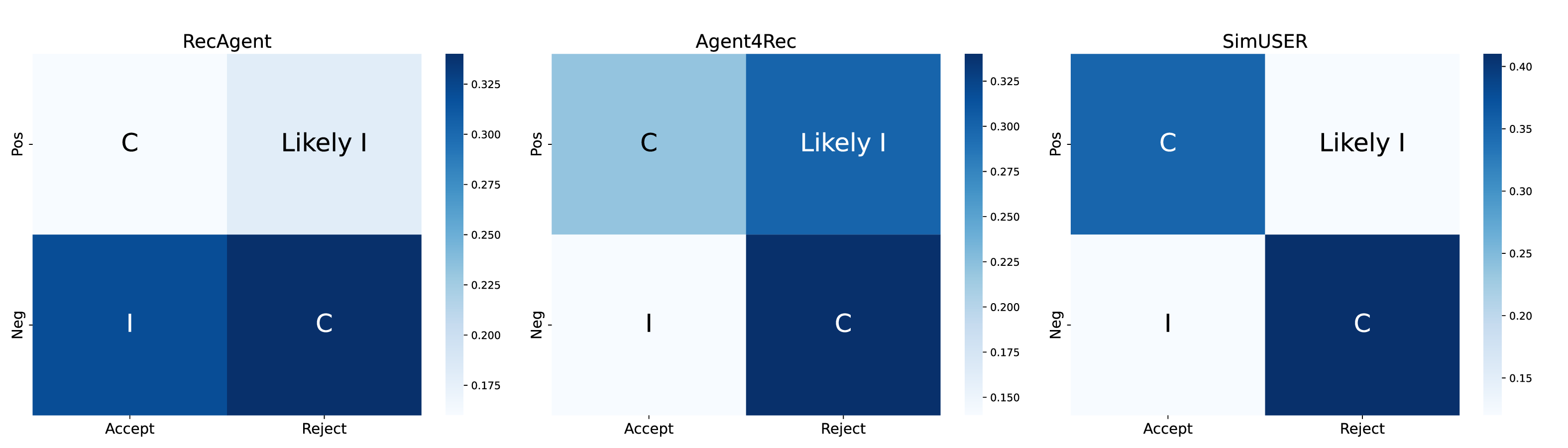} 
    \caption{Preference coherence (accept/reject task). 'I' stands for incoherent; 'C' stands for coherent. Results are reported on Tenrec dataset with hard negative items.}
    \label{fig:coherence_watch_tencent}
\end{center}
\end{figure}

To further assess the robustness of our agents under more realistic recommendation conditions, we conduct an additional experiment using the Tenrec dataset \cite{yuan2022tenrec}. Unlike the Reddit dataset, which relies on random negative sampling, Tenrec provides true negative feedback—items that were shown to users but explicitly ignored. This allows us to create harder negative samples, as these unclicked items are likely to be more relevant but still rejected by real users. Under this setting, hard negatives are items that were exposed to the user but ignored. As expected, the increased difficulty results in a slight drop in coherence across all agents (Figure \ref{fig:coherence_watch_tencent}). SimUSER remains the most consistent but sees a 5\% decrease in coherence, while Agent4Rec and RecAgent show larger declines. Notably, Agent4Rec exhibits a stronger bias toward selecting hard negatives, suggesting sensitivity to misleading but plausible recommendations.

\subsection{Impact of Persona on User Behaviors}
\begin{figure}
\captionsetup[subfloat]{farskip=2pt,captionskip=1pt}
  \centering
  \subfloat{\includegraphics[width=1.0\linewidth]{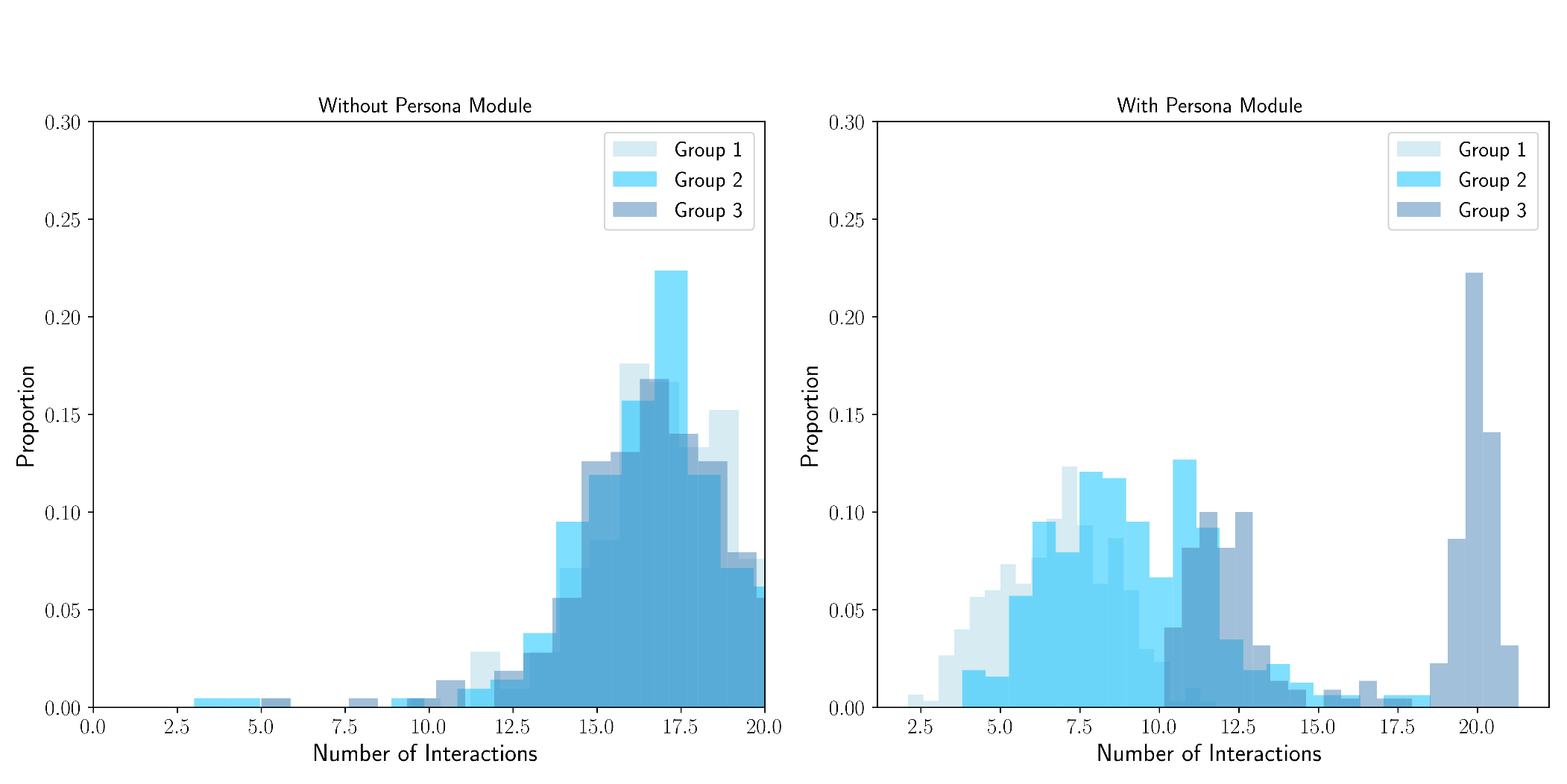}} \hfill
  \subfloat{\includegraphics[width=1.0\linewidth]{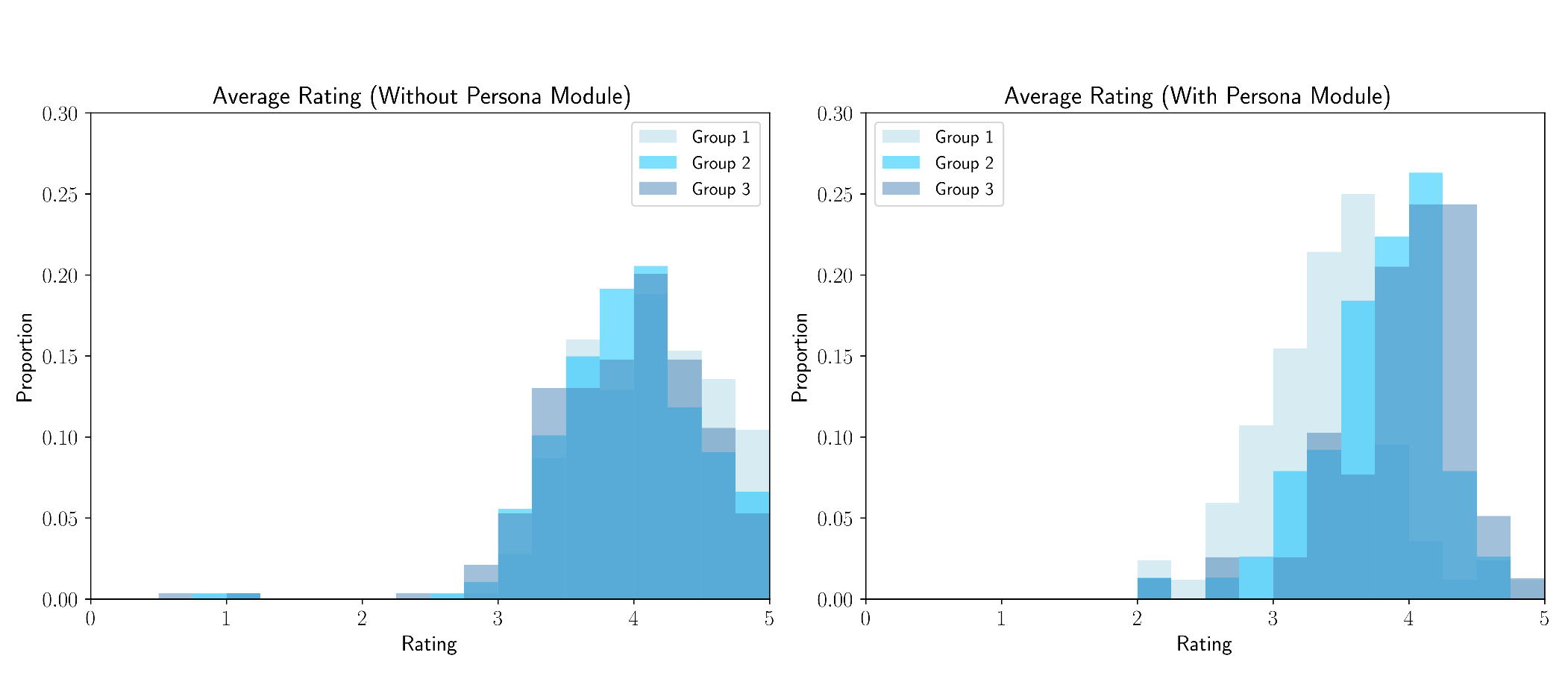}}
  \caption{Distribution of interaction numbers (top) and average ratings (bottom) for 3 groups of personas. The left column does not use persona module while the right employs a persona module.} \label{fig:impact_persona}
\end{figure}

In real life, user behaviors are driven by factors beyond mere individual tastes, including personality traits and demographic attributes such as age, and occupation. To account for these factors, we developed a persona module that incorporates these characteristics. Using the MovieLens dataset, agents were categorized based on their age, occupation (student, office worker, other), and personality traits --- categorical features were encoded using one-hot representation, via k-means (k=3). We hypothesize that agents with similar preferences may still exhibit distinct behavior patterns due to their unique persona. Figure \ref{fig:impact_persona} depicts the distribution of interaction numbers and average ratings among agents with different personas. As shown in the figure, it is evident that agents from different groups demonstrate distinct behaviors, with higher average ratings in groups 2 and 3 than group 1. Synthetic users without personas display more homogeneous behaviors. Younger users (group 3, avg age=19.6) engage more frequently with the system compared to older users (group 1, avg age=40.2), likely due to higher adaptability and interest in exploring new recommendations, while older users exhibit fewer interactions, possibly from ingrained preferences and less willingness to engage with new content.

\subsection{Post-Interaction Interviews}
\begin{figure}[htbp]
    \centering
    \begin{tcolorbox}[colframe=brown!70!black, colback=white, title=Case Study of Satisfaction Interview]
\small
\textbf{Question:} How satisfied are you with the recommender system? Please \textcolor{customgold}{rate} your satisfaction on a scale from 1 to 10 and provide an \textcolor{customgold}{explanation} for your rating.\\
\textbf{Rating:} \textcolor{customblue2}{7}\\
\textbf{Reason:} As a 35-year-old film enthusiast with a preference for independent films and documentaries, I found the recommender system fairly effective in suggesting thought-provoking content. It recommended several documentaries and indie films that I enjoyed.\\
However, the system often leaned towards mainstream and popular films rather than exploring unique and lesser-known titles. This tendency occasionally led to less satisfying recommendations, as I value discovering independent, high-quality films.\\
Additionally, while the system recognized my interest in dramas and thrillers, it missed specific sub-genres like neo-noir and psychological thrillers that I particularly enjoy. Improving the system’s ability to identify these nuanced preferences could enhance its effectiveness.\\
Overall, the system shows strengths in general recommendations but would benefit from a better balance between popular and niche films.
    \end{tcolorbox}
    \caption{One case study of post-exit feeling interview. The \textcolor{customblue2}{blue} color highlights the user's satisfaction score.}
    \label{fig:feeling_interview}
\end{figure}

A central advantage of LLM-based evaluation is their capacity to generate human-comprehensible explanations. Such explanations provide valuable insights that can be employed to refine the recommender system. Figure \ref{fig:feeling_interview} illustrates a post-interaction interview scenario, where the agent rates its experience with the recommender and returns an explanation based on its personal tastes, habits, emotional memory, and interaction history. The agent noticed that the recommender system generally aligned well with its tastes, particularly recommending movies suitable for its age and personality. However, some factors reduce the overall satisfaction. For instance, despite the agent's preference for sub-genres like neo-noir and psychological thrillers such as ``The Sixth Sense'' and ``Apt Pupil'', the system frequently recommended mainstream blockbusters like ``Star Wars: Episode I'', ``The Phantom Menace''. These findings underscore the importance of understanding the nuanced behavior of different recommendation algorithms and their suitability for each group of users.

\subsection{Rating Items under Hallucination}
\begin{figure}[tbp]
    \begin{center}
        \includegraphics[width=0.9\linewidth]{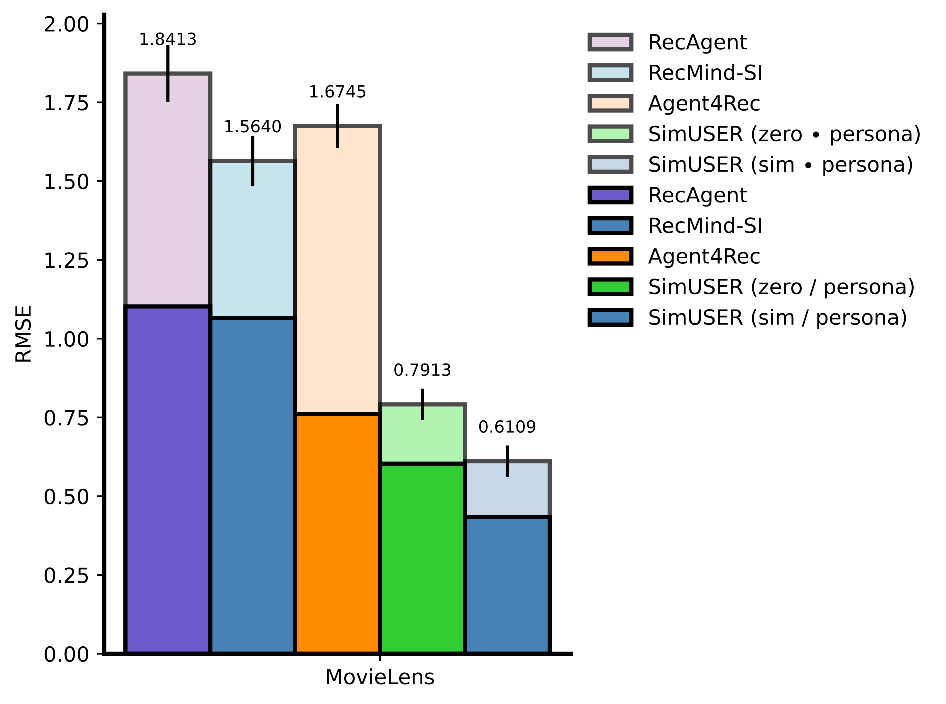} 
    \caption{Comparison of RMSE values for original (dark colors) and hallucination-affected (light colors) models for the rating task (MovieLens). }
    \label{fig:hallucination_rating}
\end{center}
\end{figure}
In this section, we specifically target items that are unfamiliar to the LLM, seeking to evaluate the ability of SimUSER to mitigate hallucination through its memory module. Similarly to Section \ref{sec:rating}, users are asked to rate movies (MovieLens). Nevertheless, we exclusively include items that are detected as unknown to the LLM. These items $i$ are identified by querying the LLM to classify each movie into one of 18 genres. If the LLM’s genre classification matches the actual category $g_{i}$, it indicates that the LLM is familiar with the item, and such movies are excluded from the experiment. From Figure \ref{fig:hallucination_rating}, it is evident that while the RMSE values for all methods increase under hallucination, the performance degradation of SimUSER is less severe compared to the baselines. This relative robustness of SimUSER can be attributed to its KG memory, which effectively mitigates the impact of hallucination by leveraging relationships between users/movies/ratings from previous interactions. By comparing the unfamiliar movie with these similar, well-known ones, the LLM can anchor its predictions in familiar contexts, reducing the likelihood of hallucinations and leading to more accurate ratings.

\subsection{Thumbnail Quality Effect}
\label{sec:eff_images}

\begin{figure}[tbp]
    \begin{center}
        \includegraphics[width=0.8\linewidth]{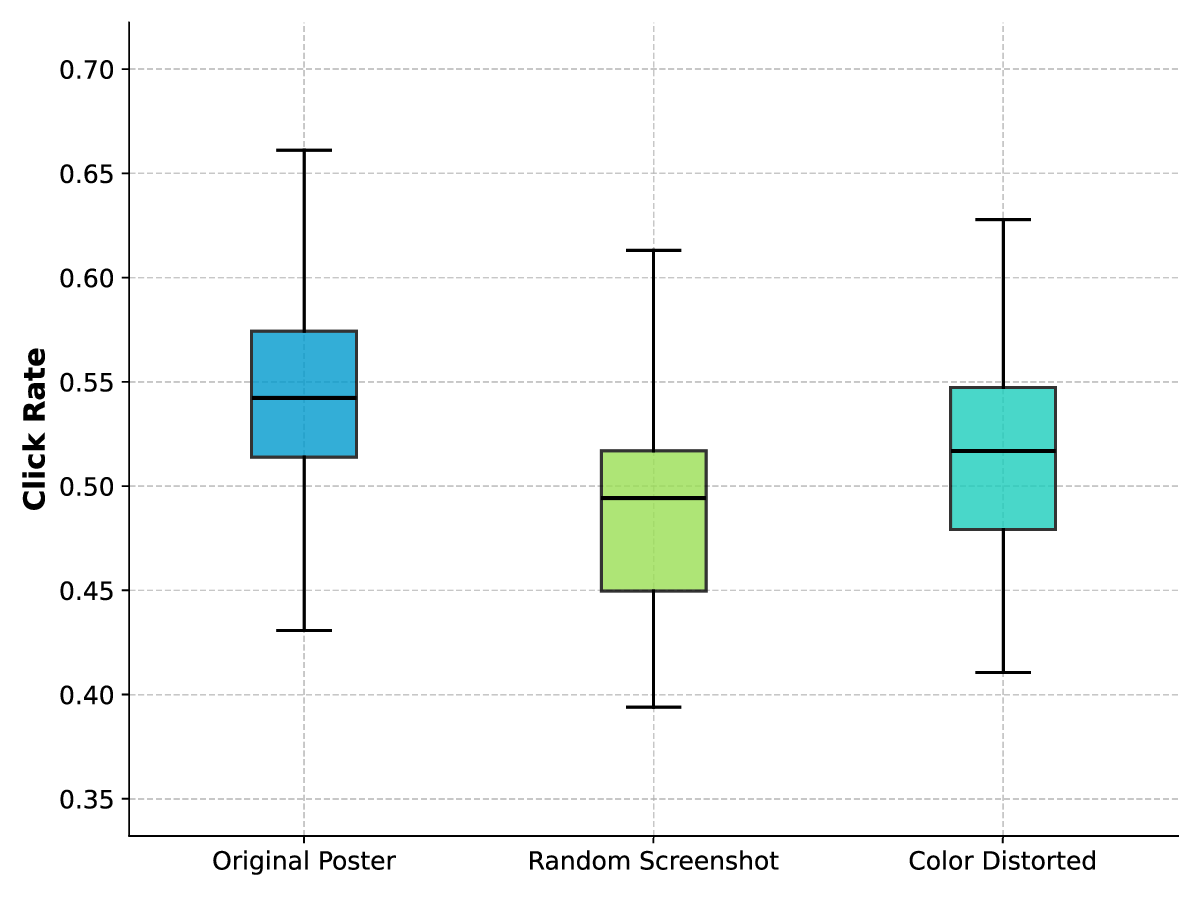} 
    \caption{Effect on visual cues on rating distribution for different thumbnail types.}
    \label{fig:rating_visual_cues}
\end{center}
\end{figure}
Emotions largely shape decision-making in the recommendation domain. At the center of emotion, images are powerful stimuli that motivate our choices. In light of this, a question arises: Can SimUSER be useful in assessing the quality of items' thumbnails? To understand the factors influencing ratings, we randomly selected 100 movies from the MovieLens dataset and ask 100 agents whether they want to watch it. For each movie, we collected three versions of the thumbnails: 1) the original movie poster, 2) a random screenshot from the movie trailer on YouTube, and 3) the original movie poster distorted with a blue color filter. Based on the click rates shown in Figure \ref{fig:rating_visual_cues}, we notice that high-quality thumbnails --- original posters, significantly influence users' inclination to watch a movie. Specifically, original posters lead to higher engagement compared to random screenshots and color-distorted posters. This result highlights SimUSER’s capability to reflect the quality of item images in decision-making processes, mirroring trends commonly observed in real-world recommender systems.

\subsection{Exposure Effect in Recommendation}

\begin{figure}[tbp]
    \begin{center}
        \includegraphics[width=1.0\linewidth]{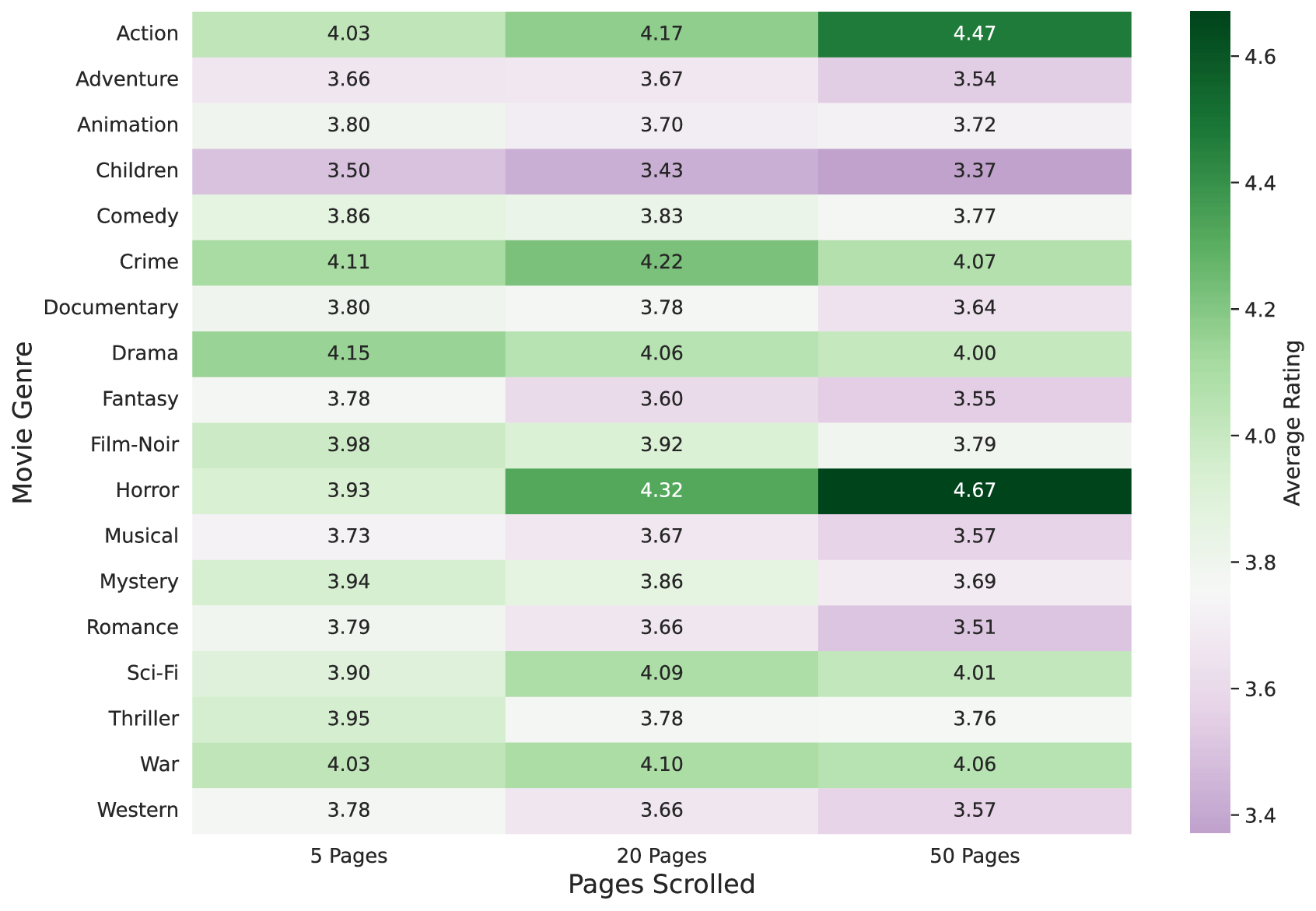} 
    \caption{Heatmap showing the impact of biased recommendations on genre ratings over time --- exposure effect. The genres and their ratings are displayed after 5, 20, and 50 pages scrolled.}
    \label{fig:exposure_effect}
\end{center}
\end{figure}

To assess how biased recommendations shape user preferences over time, we introduce a scenario where the RS only recommends two movie categories: \textit{action} and \textit{horror}. It emulates an exposure effect \cite{farber2023biases}, where repeated exposures to a particular stimulus increase an individual's preference for that stimulus. In the context of recommender systems, repeated exposure to specific genres could amplify user preferences for those genres. Under this scenario, we record the average movie ratings for each category after 5, 20, and 50 pages scrolled by the agents. Namely, the 50 agents are prompted to rate 500 randomly selected movies. Figure \ref{fig:exposure_effect} illustrates a tendency of the agents to rate items of categories that are over-represented higher during the interactions with the recommender system, particularly after more than 20 pages. Conversely, categories that differ significantly from \textit{action} and \textit{horror} genres generally tend to receive lower average ratings. Experimental results validate SimUSER's capability to replicate the exposure effect, although further research and validation are required with alternative datasets.

\subsection{User Review Influence}
\begin{table}[tbp]
\centering
\resizebox{1.0\columnwidth}{!}{
\begin{tabular}{lcccccc}
\toprule
 & \multicolumn{2}{c}{\textbf{MF}} & \multicolumn{2}{c}{\textbf{MultVAE}} & \multicolumn{2}{c}{\textbf{LightGCN}} \\
\cmidrule(r){2-3} \cmidrule(r){4-5} \cmidrule(r){6-7}
\textbf{Condition} & $\overline{P}_{\text{view}}$ & $\overline{P}_{\text{like}}$ & $\overline{P}_{\text{view}}$ & $\overline{P}_{\text{like}}$ & $\overline{P}_{\text{view}}$ & $\overline{P}_{\text{like}}$ \\
\midrule
\rowcolor{blue!10}
Origin & 0.461 & 0.443 & 0.514 & 0.455 & 0.557 & 0.448 \\
\phantom{  }+ With \# Reviews & 0.485 & 0.491 & 0.535 & 0.492 & 0.570 & 0.505 \\
\phantom{  }+ With Negative & 0.413 & 0.408 & 0.450 & 0.435 & 0.507 & 0.409 \\
\phantom{  }+ With Positive & 0.469 & 0.495 & 0.549 & 0.510 & 0.573 & 0.444 \\
\bottomrule
\end{tabular}}
\caption{Impact of user reviews on recommender System performance.}
\label{tab:review_impact}
\end{table}
User proxies may help researchers in identifying the psychological effect of reviews on human interactions. To investigate this effect, we modified the recommendation simulator to display a) the number of reviews, b) one random negative comment, or c) one random positive comment for each item on the recommendation page. We report in Table \ref{tab:review_impact} the average viewing ratio $\overline{P}_{\text{view}}$ and ratio of likes $\overline{P}_{\text{like}}$. We can see that displaying the number of reviews slightly improves the viewing ratio, especially for items having enough reviews (i.e., more than 20 reviews). This aligns with humans' inclination to select popular items in real-life scenarios. On the other, there is no significant difference in $\overline{P}_{\text{like}}$ (t-test $p>0.05$). Another observation is that displaying negative reviews has a stronger impact on user behavior than showing positive reviews, with a decrease in both the average viewing ratio and number of likes. One possible explanation is that negative reviews discourage users from watching an item, while positive reviews primarily encourage users who are already inclined to watch it to proceed with their choice.   

\subsection{SimUSER vs. Offline Metrics}

\begin{table}[btp]
\centering
\resizebox{1.0\columnwidth}{!}{
\begin{tabular}{lcccc}
\toprule
 & \multicolumn{2}{c}{nDCG@10} & \multicolumn{2}{c}{F1-score@10} \\
\cmidrule(r){2-3} \cmidrule(r){4-5}
Method & Offline & SimUSER & Offline & SimUSER \\
\midrule
MF & 0.226 & 0.213 & 0.165 & 0.144 \\
MultVAE & 0.288 & 0.278 & 0.180 & 0.189 \\
LightGCN & \textbf{0.423}  & \textbf{0.465} & \textbf{0.227} & \textbf{0.255} \\
\bottomrule
\end{tabular}}
\caption{nDCG@k (k=10) and F1-score@k (k=10) for three recommender systems, using either offline or SimUSER-generated interactions.}
\label{tab:ndcg_recall}
\end{table}

We aim to investigate whether SimUSER can reliably estimate traditional metrics such as nDCG@k (k=10) and F1-score@k (k=10) by comparing the results from traditional offline evaluation with those from SimUSER-generated interactions. For this purpose, we evaluate three recommender systems using the MovieLens dataset under identical conditions for both offline and SimUSER-based evaluations. Table \ref{tab:ndcg_recall} reports the nDCG@k and F1-score@k ($k$=10) for both evaluation strategies. In the SimUSER scenario, interactions are generated by our synthetic users, where liked and disliked items replace the ground-truth interactions from the offline dataset. Results indicate minimal differences between SimUSER-generated and real-world data, with consistent model rankings across systems. These slight differences reflect real-world users being unconstrained by page numbers and interaction frequency. These findings demonstrate that SimUSER reliably measures traditional metrics while enabling exploration of system performance across user demographics, website settings (items per page), and recommender system configurations.

\subsection{Impact of Number of Interactions on Rating Performance}

\begin{figure}[tbp]
    \begin{center}
        \includegraphics[width=1.0\linewidth]{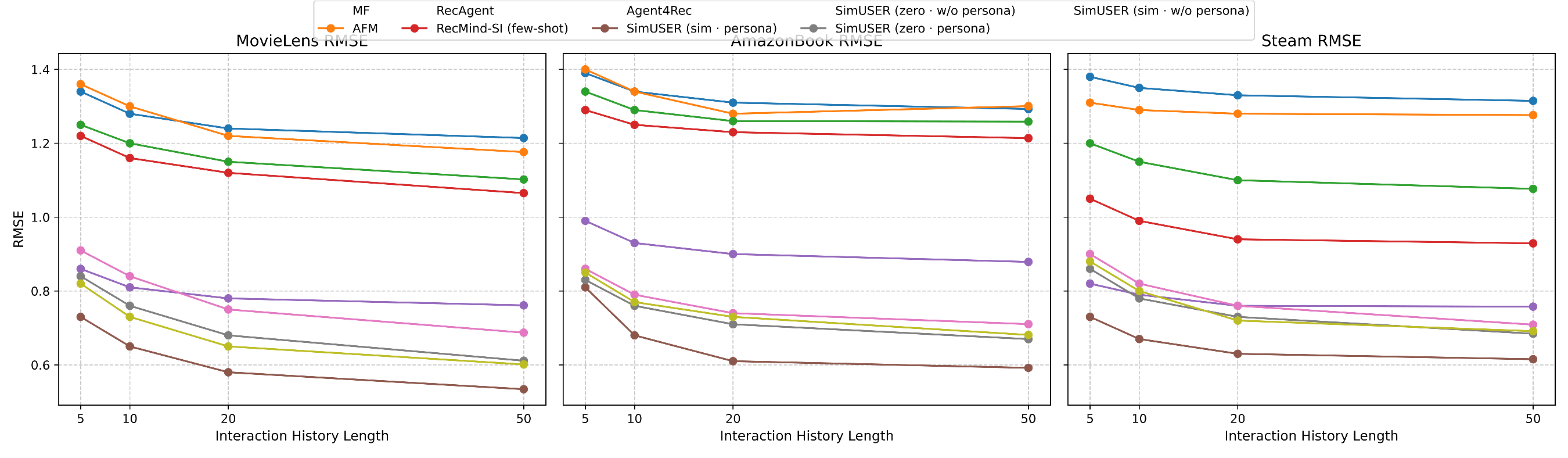} 
    \caption{Impact of history size on rating prediction performance (RMSE) across datasets.}
    \label{fig:history_length}
\end{center}
\end{figure}

In this experiment, we measure rating prediction performance as a function of interaction history length $\in$(5, 10, 20, and 50 interactions). While most methods generally benefit from increased context (Figure \ref{fig:history_length}), small fluctuations occur (e.g., AFM on AmazonBook shows a slight rise from 1.28 at 20 interactions to 1.3006 at 50). SimUSER consistently outperforms all baselines, achieving RMSEs of 0.5020 (MovieLens), 0.5676 (AmazonBook), and 0.5866 (Steam) at 50 interactions. These results confirm that leveraging persona-based context yields robust performance improvements, even with limited historical data, and aligns with our main results. This highlights SimUSER's ability to utilize past interactions for realistic simulations while remaining believable when modeling \textit{cold-start} or \textit{few-shot} users.

\subsection{Ablation Studies}

\subsubsection{Impact of the Knowledge-Graph Memory on SimUSER}

\begin{table*}[tbp]
\centering
\begin{tabular}{lcccccc}
\toprule
\textbf{Methods} & \multicolumn{2}{c}{\textbf{MovieLens}} & \multicolumn{2}{c}{\textbf{AmazonBook}} & \multicolumn{2}{c}{\textbf{Steam}} \\
 & \textbf{RMSE} & \textbf{MAE} & \textbf{RMSE} & \textbf{MAE} & \textbf{RMSE} & \textbf{MAE} \\
\midrule
\rowcolor{blue!10}
SimUSER(zero) \textcolor{persona_color}{$\varheartsuit$}  & \underline{0.5813}* & \underline{0.5298}* & 0.6542 & 0.5116* & 0.6798* & 0.6151* \\
\phantom{  } SimUSER(zero) \textcolor{no_persona_color}{$\clubsuit$}  & 0.6545 & 0.6299 & 0.6771 & 0.6210 & 0.7176 & 0.6533 \\
\midrule
\rowcolor{blue!10}
SimUSER(sim) \textcolor{persona_color}{$\varheartsuit$} & \textbf{0.5020}* & \textbf{0.4465}* & \textbf{0.5676}* & \textbf{0.4210}* & \textbf{0.5866}* & \textbf{0.5325}* \\
\phantom{  } SimUSER(sim) \textcolor{no_persona_color}{$\clubsuit$} & 0.6300 & 0.6336 & \underline{0.6109} & \underline{0.4881} & \underline{0.6482} & 0.6481 \\
\bottomrule
\end{tabular}
\caption{Performance comparison in rating prediction for agents equipped with (top two rows \textcolor{persona_color}{$\varheartsuit$}) and without a KG memory (bottom two rows \textcolor{no_persona_color}{$\clubsuit$}). Asterisks (*) denote statistically significant improvements when the KG memory is used.}
\label{fig:impact_knowledge_graph}
\end{table*}
Here, our focus is on evaluating the impact of incorporating a knowledge-graph memory on the performance. Specifically, the goal is to determine whether employing the KG memory, which simulates external influences such as reviews, enhances believability in human proxies. All models follow the same settings as in Sec \ref{sec:rating}. Table \ref{fig:impact_knowledge_graph}, highlights that leveraging the KG structure significantly reduces both RMSE and MAE across different datasets. This module mirrors how our prior expectations of an item can shape and bias our assessment of it.

\subsubsection{Persona Matching: Age, Occupation}

\begin{table}[btp]
\centering
\begin{tabular}{lcc}
\toprule
\textbf{Metric} & \textbf{Age} & \textbf{Occupation} \\ 
\midrule
Accuracy & 0.7230 & 0.6764 \\ 
Precision & 0.7586 & 0.6933 \\ 
Recall & 0.7921 & 0.7430 \\ 
F1 Score & 0.7749 & 0.7172 \\ 
\bottomrule
\end{tabular}
\caption{Performance of Persona Matching in Predicting Age and Occupation Using the MovieLens-1M Dataset.}
\label{table:persona_matching_results}
\end{table}
\begin{figure}[tbp]
    \begin{center}
        \includegraphics[width=0.70\linewidth]{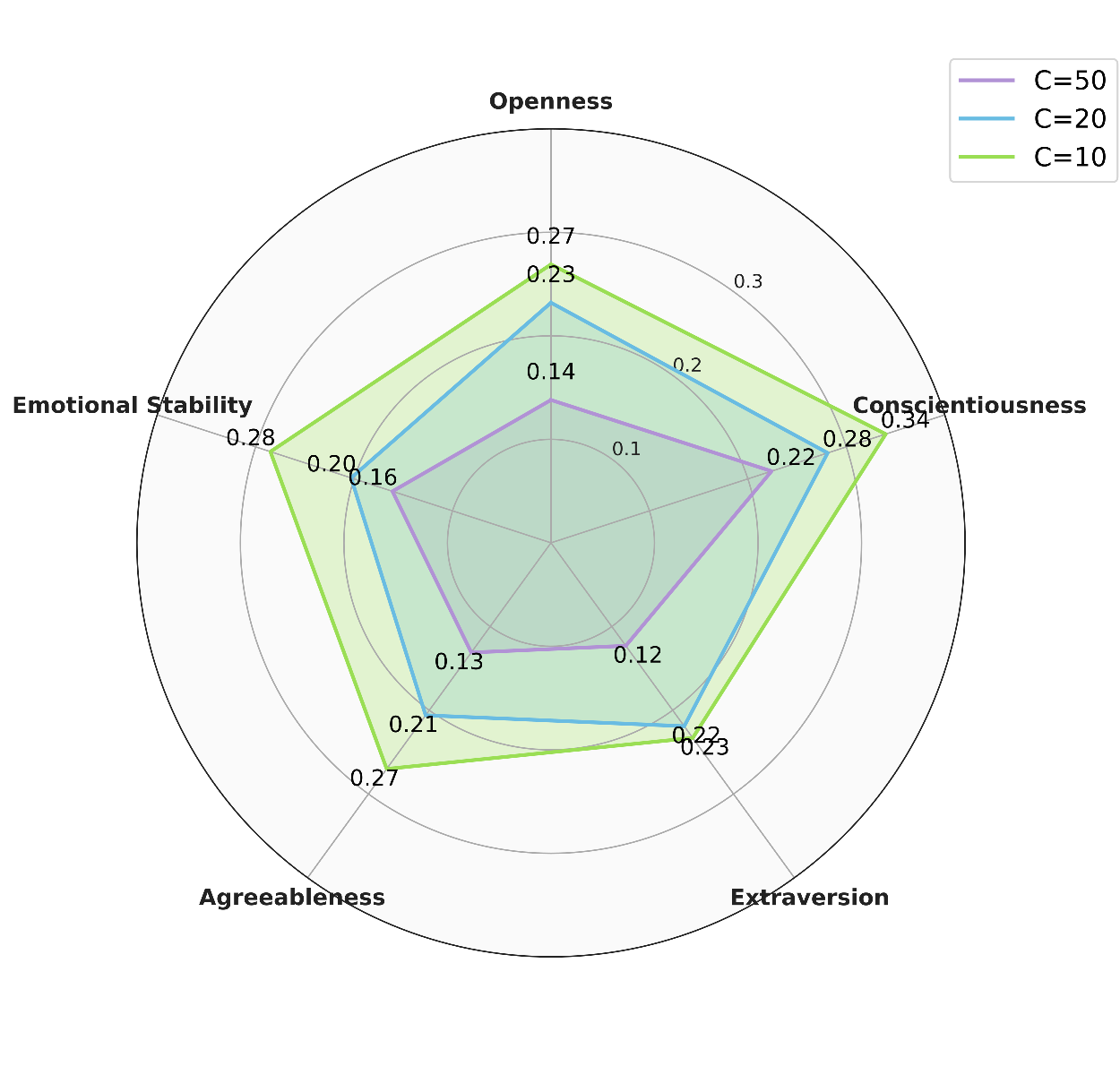} 
    \caption{MAE of personality trait predictions for different C values.}
    \label{fig:personality_matching}
\end{center}
\end{figure}

We postulate that personas are crucial for capturing the heterogeneity and diversity present in real-world recommender networks. These attributes significantly shape individual behaviors and preferences, which subsequently influence the overall dynamics of the system. To evaluate the effectiveness of our self-consistent persona-matching technique, we conducted an experiment using the MovieLens-1M dataset. The goal was to predict the age and occupation of users based on their historical data. This task was formulated as a classification problem. Our results are summarized in table \ref{table:persona_matching_results}. We observe a high degree of alignment between the predicted and actual user personas, highlighting the effectiveness of Phase 1 in SimUSER. Overall, \textit{persona matching} turns out to be reasonably robust for enriching simulated agents with detailed backgrounds, including domains where explicit demographic data is not readily provided.

\subsubsection{Persona Matching: Personality}
In order to assess the quality of persona matching in predicting personality traits from historical interaction data, we conduct an additional experiment using the Personality 2018 dataset \cite{nguyen2018user}. The primary objective is to evaluate whether our model could accurately infer users' Big Five personality traits based solely on users' watching history. For a fair comparison, the personality traits within the dataset, as well as the predictions, are normalized to a scale ranging from 0 to 1. We report the results for various lengths of movie history $\varrho \in \{10, 20, 50\}$. This task is formulated as a regression problem. Figure \ref{fig:personality_matching} summarizes the results, showing that our model achieved an average MAE of $0.155$ across all traits. Besides, the results reveal that using a history length of 50 items reduces the average MAE from $0.279$ (10 items) to $0.155$, demonstrating that self-consistent persona matching can reasonably predict personality traits of users from their past interactions.

\subsubsection{Choice of Foundation Model}

\begin{table*}[tbp]
\centering
\begin{tabular}{lcccccc}
\toprule
\textbf{Methods} & \multicolumn{2}{c}{\textbf{MovieLens}} & \multicolumn{2}{c}{\textbf{AmazonBook}} & \multicolumn{2}{c}{\textbf{Steam}} \\
 & \textbf{RMSE} & \textbf{MAE} & \textbf{RMSE} & \textbf{MAE} & \textbf{RMSE} & \textbf{MAE} \\
\midrule
GPT-4o-mini & \underline{0.5020} & \underline{0.4465} & \underline{0.5676} &  \underline{0.4210} & \underline{0.5866} & \underline{0.5325} \\
GPT-4o & \textbf{0.4739} & \textbf{0.4167} & \textbf{0.5532} & \textbf{0.3998} & \textbf{0.5549} & \textbf{0.4823} \\
Mistral-7b Instruct & 0.5486 & 0.4874 & 0.6435 & 0.4909  & 0.6407 & 0.6275 \\
Llama-3 Instruct & 0.5901 & 0.5812 & 0.6346 & 0.4715 & 0.6453 & 0.6321 \\
Phi-3-mini & 0.6358 & 0.5964 & 0.6789 & 0.5763  & 0.7175 & 0.6935 \\
\bottomrule
\end{tabular}
\caption{Performance comparison in rating prediction on MovieLens with different types of foundation LLMs.}
\label{fig:foundation_model}
\end{table*}

We seek to evaluate the performance of our methodology using various foundation models on the movie rating task. Specifically, we compare the results obtained by employing GPT-4o-mini, GPT-4o, Mistral-7b Instruct, Llama-3 Instruct, and Phi-3-mini as the underlying LLMs. The results, presented in Table \ref{fig:foundation_model}, demonstrate that the performance of SimUSER is generally robust across different foundation models. While GPT-4o exhibits significantly lower mean RMSE and MAE (t-test $p<0.05$), GPT-4o-mini achieves similar performance but with a lower inference time. Mistral-7b Instruct also performs reasonably well on the MovieLens dataset. On the other hand, Llama-3 Instruct and Phi-3-mini, while competitive, show higher errors.


\subsubsection{Impact of Perception Module}
\definecolor{myGreen}{RGB}{11,84,0}
\definecolor{myblue}{RGB}{2,125,181}
\definecolor{myRed}{RGB}{255,123,122}

We now investigate the perception module's impact on agent believability. Table \ref{tab:perception_module} shows agents consistently exhibit more realistic behavior with the perception module (\textcolor{myblue}{$\spadesuit$}), likely due to the inclusion of visual details and unique selling points. The believability gain is lower on AmazonBook than other datasets, possibly because users judge books less by covers and more by descriptions. Examining interactions reveals agents with different personas are significantly influenced by emotional tones. For instance, an agent with high openness may be more inclined to select movies with captions that use positive language like ``exciting'' or ``inspiring''. While SimUSER (\textcolor{myRed}{$\vardiamondsuit$}) may inherit biases from the LLM's interpretation of item descriptions, these can be partially mitigated through factual caption information. This suggests the perception module contributes to more visually and emotionally driven engagement.

\begin{table}[btp]
\centering
\resizebox{1.0\columnwidth}{!}{
\begin{tabular}{lcccc}
\toprule
 & \textbf{MovieLens} & \textbf{AmazonBook} & \textbf{Steam} \\
\midrule
RecAgent & 3.01 $\pm$ 0.14 & 3.14 $\pm$ 0.13 & 2.96 $\pm$ 0.17   \\
Agent4Rec & 3.04 $\pm$ 0.12 & 3.21 $\pm$ 0.14 & 3.09 $\pm$ 0.16   \\
\rowcolor{blue!10}
SimUSER (\textcolor{myRed}{$\vardiamondsuit$}) & \underline{4.27$\pm$0.18} & \underline{3.94$\pm$0.16} & \underline{3.89$\pm$0.20} \\
\rowcolor{blue!10}
SimUSER (\textcolor{myblue}{$\spadesuit$}) & \textbf{4.41$\pm$0.16}* & \textbf{3.99$\pm$0.18}* & \textbf{4.02$\pm$0.23}*  \\
\bottomrule
\end{tabular}}
\caption{Human-likeness score evaluated by GPT-4o for SimUSER without (\textcolor{myRed}{$\vardiamondsuit$}) and with (\textcolor{myblue}{$\spadesuit$}) perception module. Asterisks (*) denote statistically significant improvements when the perception module is activated.}
\label{tab:perception_module}
\end{table}

\section{Discussion}
We acknowledge that our method has certain limitations. Observed behaviors are well-aligned with existing theories and common behaviors in the recommendation domain. Phenomena at micro-level (rating, watching) are manifestations of agent endogenous behaviors. But why agents possess these behaviors are unexplored due to the black-box nature of the large language models we adopted. A potential reason could be that LLMs are trained on a massive corpus that includes texts from various domains.

A potential limitation of our approach lies in its reliance on sufficient interaction data to construct detailed user personas. In some scenarios, many users exhibit limited engagement, particularly in cold-start settings where new users have few or no recorded interactions. This constraint reduces the effectiveness of our persona module, as it derives user preferences primarily from past interactions.To address this issue, a potential alternative is initializing the persona module using predefined user features, such as categorical tags (e.g., "tech-savvy," "frequent traveler").

LLMs may replicate biases prevalent in social spaces, such as some groups of individuals being underrepresented. This is problematic if it causes designers to then underlook these peoples' needs when designing a recommender system. In our experiments, we mitigated this risk by ensuring a broad range of personas via diverse potential occupations, age, and personalities. We also measured the discrepancy between identified and real personas. Our future investigation will focus on analyzing underrepresented user groups, as well as evaluating persona matching on a wider range of domains (e.g., food).

Finally, UX and UI drive our choices and actions in real-world applications. Our simulation, on the other hand, does not fully replicate all those intricate factors, which introduces a gap between real life and simulation. An important future direction is developing an image-based simulator to better capture the complex nature of user experience.

\section{Cost Analysis}
We report the cost of running SimUSER per 1000 users. Costs may vary slightly due to differences in interaction numbers and LLM outputs, but scale approximately linearly with user count. Our implementation uses OpenAI's GPT-4o-mini. SimUSER costs approximately \$13 (\$0.0013/User), while Agent4Rec costs approximately \$10 (\$0.0010/User). The cost difference mainly stems from the integration of images to enable visual-driven reasoning.

\section{Running Time Analysis}

We compare the running time of SimUSER and Agent4Rec for 1,000 user interactions with GPT-4o. Without parallelization (\textcolor{myRed}{$\varheartsuit$}), Agent4Rec and SimUSER require 9.3h and 10.1h, respectively. With parallelization (\textcolor{myblue}{$\clubsuit$}, max 500 workers), these times drop to 0.53h and 0.59h. This demonstrates that parallelizing LLM calls significantly reduces inference time, allowing the system to scale efficiently.


\end{document}